\documentclass[12pt]{article}
\usepackage[cp1251]{inputenc}
\usepackage[russian]{babel}
\usepackage{color}
\usepackage{amssymb}
\usepackage{latexsym}
\usepackage{enumerate}
\usepackage{amsmath}
\usepackage{graphicx}

\def\adimH{\mathsf{K}} % Размерность квантового гильбертова пространства
\def\aNumbers{\mathcal{N}} % Какое нибудь множество чисел
\def\bmat{\begin{bmatrix}}
\def\emat{\end{bmatrix}}
\def\C{\mathbb{C}} % Complex numbers
\def\classN{\mathsf{m}} % Число классов
\def\e{\mathrm{e}}
\def\Hspace{\mathcal{H}} % Hilbert space
\def\idmat{\mathrm{I}} % Unit matrix:\mathbb{I}\mathbf{I}\mathrm{\mathbf{I}}
\def\ig{\gamma} % Internal group element
\def\iG{\Gamma} % Internal group
\def\iGN{{\cabs{\iG}}} % Internal group size
\def\iGX{\iG^{\X}} % Exponential set of group-valued functions on space
\def\ls{\sigma} % Element of set of local state
\def\lS{\Sigma} % Set of local states
\def\lSN{{\cabs{\lS}}} % Size of set of local states
\def\lSX{\lS^{\X}} % Exponential set of functions on space
\def\id{\mathbf{1}} % Group identity:\mathsf{1}\mathbf{1}\mathbf{I}{1\!\!\mathrm{I}}
\def\N{\mathbb{N}}
\def\Nat{\N} % Натуральные числа с нулём {\N_0}
\def\onbeq{\alpha} % Symbol of orthonormal basis element of Hilbert space of quantum representation
\def\Partransport{\rho}
\def\R{\mathbb{R}} % Вещественные числа
\def\regrep{\mathrm{P}} % Регулярное представление
\def\Repbare{\rho}
\def\repirr{D} % Irreducible component symbol \Delta
\def\runisymb{\mathsf{r}} % Символ примитивного корня из единицы без указания экспоненты
\def\sg{\mathsf{f}} % Space group element
\def\sG{\mathsf{F}} % Space group
\def\sGN{{\cabs{\sG}}} % Space group size
\def\tin{0} % Начало эволюции
\def\tfin{T} % Конец эволюции
\def\Time{\mathcal{T}} % Time
\def\transmatr{\mathrm{T}} % Basis transforming matrix
\def\wg{\mathsf{g}} % Element of whole symmetry group: \mathrm{g}\mathsf{w}\mathfrak{w}
\def\wG{\mathsf{G}} % Whole symmetry group: \mathcal{W}\mathrm{G}
\def\wGN{\mathsf{M}} % Size of whole group: {N_\wG}
\def\ws{\mathsf{s}} % Element of whole set of states:\mathrm{s}
\def\wS{\mathsf{S}} % Whole set of states:\mathrm{S}
\def\wSN{\mathsf{N}} % Size of whole set of states:{N_\wS}
\def\x{\mathsf{x}} % Point of space
\def\X{\mathsf{X}} % Set of space points
\def\XN{{\cabs{{\X}}}} % Number of space points
\def\Z{\mathbb{Z}}
\newcommand{\Aut}[1]{\mathrm{Aut}\vect{#1}} % Автоморфизм
\newcommand{\barket}[1]{\left|#1\right\rangle} % |#1>
\newcommand{\brabar}[1]{\left\langle#1\right|} % <#1|
\newcommand{\cabs}[1]{\left|#1\right|} % |#1|
\newcommand{\cconj}[1]{\overline{#1}} % Complex conjugate
\newcommand{\class}[1]{K_{#1}} % Conjugate class C_{#1}
\newcommand{\CyclG}[1]{\mathsf{C}_{#1}} % Циклическая группа
\newcommand{\DihG}[1]{\mathsf{D}_{#1}} % Диэдральная группа
\newcommand{\Grassnorm}[1]{\left\|#1\right\|} % ||#1||
\newcommand{\inner}[2]{\left\langle#1\mid#2\right\rangle} % <#1|#2>
\newcommand{\innerstandard}[2]{\left(#1\mid#2\right)} % (#1|#2)
\newcommand{\Math}[1]{$#1$} % Математическое выражение внутри строки текста
\newcommand{\Mathh}[1]{$$#1$$} % Математическое выражение отдельной строкой
\newcommand{\Mone}[1]{\bmat#1\emat} % 1-компонентный вектор (матрица)
\newcommand{\Mthree}[9]{\bmat#1&#2&#3\\
 #4&#5&#6\\
 #7&#8&#9\emat} % 3x3 матрица
\newcommand{\Mtwo}[4]{\bmat#1&#2\\#3&#4\emat} % 2x2 matrix
\newcommand{\ordset}[1]{\left[#1\right]} % Set of elements
\newcommand{\Perm}[1]{\mathrm{Sym}\left(#1\right)} % Permutation group
\newcommand{\ProbBorn}[2]{\Prob\!\vect{#1,#2}} % Борновская вероятность
\newcommand{\Rep}[1]{\Repbare\left(#1\right)} % Представление на элементе группы
\newcommand{\runi}[1]{\runisymb_{#1}} % e^(2 Pi i/n) главный корень из единицы \zeta \epsilon
\newcommand{\set}[1]{\left\{#1\right\}} % Set of elements
\newcommand{\SL}[2]{\mathsf{SL}\vect{#1,#2}} % Специальная линейная группа SL(n,q)
\newcommand{\SymG}[1]{\Perm{#1}} % Симметрическая группа
\newcommand{\vect}[1]{\left(#1\right)} % Ordered set of elements, round brackets
\newcommand{\Vthree}[3]{\bmat#1\\#2\\#3\emat} % 3-component vector
\newcommand{\Vtwo}[2]{\bmat#1\\#2\emat} % 2-компонентный вектор

\begin{document}
%\setcounter{page}{320}
%---English part$$$$$$$$$$$$$$$$$$$$$$$$$$$$$$$$$$$$$$$$$$$$$$$$$$$$$$$$
\begin{center}
{\Large\bf
%Конечные квантовые модели: конструктивный подход к описанию квантового поведения
Finite Quantum Models: Constructive Approach\\ to Description of Quantum Behavior
} 
\vskip 10pt 
%\copyright~~
{\large\bf 
%2010 
{Vladimir V. Kornyak}}\\[10pt]
\emph{Laboratory of Information Technologies}, %\\
Joint Institute for Nuclear Research\\
\emph{141980 Dubna, Moscow region, Russia}\\
\emph{E-mail: kornyak@jinr.ru}
\end{center}
%\vskip 10pt
%{\small
%\noindent
\def\abstractname{Abstract}
\begin{abstract}
Universality of quantum mechanics --- its applicability to physical systems of quite
different nature and scales --- indicates that quantum behavior can be a manifestation 
of general mathematical properties of systems containing indistinguishable, i.e. lying
on the same orbit of some symmetry group, elements. 
In this paper we demonstrate, that quantum behavior arises naturally in systems 
with finite number of elements connected by nontrivial symmetry groups.
The ``finite'' approach allows to see the peculiarities of quantum description more distinctly
without need for concepts like ``wave function collapse'', ``Everett's multiverses'' etc.
In particular, under the finiteness assumption any quantum dynamics is reduced to 
the simple permutation dynamics.
The advantage of the finite quantum models is that they can be studied constructively
by means of computer algebra and computational group theory methods.
\end{abstract}

%}
\section{Introduction}
The question of ``whether the real world is discrete or continuous'' or even ``finite 
or infinite'' is entirely \emph{metaphysical}, since neither empirical observations nor
logical arguments can validate one of the two adoptions --- this is a matter of belief or
taste. Of course, discrete and continuous \emph{mathematical} theories differ essentially
and effectiveness of their applications in physics depends on specific historical
background. In particular, since Newton's time to advent of modern computers analysis
and differential geometry were, in fact, the only tools for mathematical study of 
physical systems% 
\footnote{Poincar\'{e} emphasized conventionality of choice between discrete (finite)
and continuous (infinite) descriptions of nature. It is interesting to trace
evolution of his personal preferences.
%\par
In the book ``Value of science''
(\cite{PoincareValEn}, pp. 80--81 of English translation) he, denying fundamental 
validity of the concept of continuum, appreciates its heuristic power very much:
``\emph{The sole natural object of mathematical thought is the whole 
number. It is the external world which has imposed the continuum 
upon us, which we doubtless have invented, but which it has forced us 
to invent. Without it there would be no infinitesimal analysis; all 
mathematical science would reduce itself to \textbf{arithmetic} or to the 
\textbf{theory of substitutions}. On the contrary, we have devoted to the study 
of the continuum almost all our time and all our strength. 
\textellipsis{}Doubtless it will be said that outside of the whole number
 there is no rigor, and consequently no mathematical truth; that the whole 
number hides everywhere, and that we must strive to render 
transparent the screens which cloak it, even if to do so we must resign 
ourselves to interminable repetitions. Let us not be such purists and 
let us be grateful to the continuum, which, if \textbf{all} springs from the 
whole number, was alone capable of making \textbf{so much} proceed there from.}'' 
\par
Several years later
 --- after the first observations of quantum behavior in physical
systems --- Poincar\'{e} writes (\cite{PoincareNewEn}, p. 643 of Russian translation):
``\emph{Now we can not say that
``nature does not make jumps'' (Natura non facit saltus), --- 
in fact, it acts just the other way around.
And not only is the matter reduced, possibly, to atoms,
but even the world history and, I would say, the time itself,
since two instants within the interval between two leaps can not be distinguished
because they belong to the same state of the universe.}''
%\par
Then Poincar\'{e} resumes his view of the problem:
``\emph{But we should not be too hasty, because the only, that is obvious 
now, is that we are quite far from the end of the struggle between 
two styles of thought --- one, typical for atomists believing
in existence of elementary items, such that very large, but finite, number
of their combinations is enough for explanation of the whole variety
of aspects of the universe, and another, inherent to the followers
of ideas of continuity and infinity.}''}.
%------------------------------------------------------------------------------
With the development of digital technologies the ``discrete'' style of thought
becomes more and more popular, and the real abilities of discrete mathematics
in applications have increased substantially.
An important advantage of discrete description is its conceptual ``economy'' in
Occam's sense --- the absence of ``superfluous entities'' based on the idea of actual 
infinity such as ``Dedekind cuts'', ``Cauchy sequences'' etc.
Moreover, in fact, discrete mathematics is richer than continuous
 --- continuity ``smooths'' subtle details of structures.
To realize this thesis, compare the lists of simple Lie groups and simple
finite groups.
There are also many arguments that discrete description of physical processes
at small (Planck) distances is more adequate and that it makes sense to treat continuity
only as a logical framework for approximate (``thermodynamic'') description of large
collections of discrete structures.
\par
A remarkable feature of quantum mechanics is its universality.
It is suitable for description of systems of quite different physical nature 
in a wide range of sizes --- from elementary particles to large
molecules. For example, experimental observations of quantum-mechanical interference 
between the fullerene \Math{C_{60}} molecules are described in \cite{fullerinterferEn}.
Such universality is usually inherent in theories at the heart of which some 
\emph{a priori} mathematical principles lie.
An example of such universal mathematical scheme is \emph{statistical mechanics}. 
It is based on independent of specific physical system principles. Most important 
of them are classification of microstates in accordance with their energies and 
postulate of equipartition of energy over degrees of freedom.
In the case of quantum mechanics, the leading mathematical principle is symmetry.
Only systems containing indistinguishable particles demonstrate quantum-mechanical
behavior --- any violation of identity of particles destroys quantum interferences.
\par
In this paper we consider the main constituents of quantum description under 
the assumption of finiteness of all sets involved in the constructions.
With this approach ``everything can be reduced to arithmetic and to 
the theory of substitutions'', in the words of Poincar\'{e}.
%_____________________________________________________________________
\section{Basic Constructions and Notations}
%_____________________________________________________________________
\subsection{Classical and quantum evolution}
We consider evolution of \emph{dynamical system} 
with the finite set of \emph{states}
\Math{\wS=\set{\ws_1,\ldots,\ws_{\wSN}}}
in the \emph{discrete time}
\Math{t\in\Time=\set{\ldots,-1,0,1,\ldots}}.
Considering \emph{finite} evolutions we can assume, 
for simplicity of notation, that
\Math{\Time=\ordset{\tin,1,\ldots,\tfin}},
where \Math{\tfin\in\Nat}. 
\par
\emph{Classical evolution} (or \emph{history}, or \emph{trajectory}) of the dynamical 
system is a sequence of states depending on time
\Math{\ldots,s_{t-1},s_{t},s_{t+1},\ldots\in\wS^\Time}.
\par
We assume that a finite \emph{symmetry group}
 \Math{\wG=\set{\wg_1=\id,\ldots,\wg_{\wGN}}}
acts on the set of states: \Math{\wG\leq\Perm{\wS}}.
\par
A sequence of permutations 
\Math{\ldots{}p_{t-1},p_{t},p_{t+1}\ldots\in\wG^\Time},
determining the product
\Math{\cdots{}p_{t-1}\,p_{t}\,p_{t+1}\cdots\in\wG},
will be called \emph{quantum evolution} --- 
the meaning of this definition will be clarified in what follows.
%______________________________________________________
\subsection{Dynamical systems with space structure}
In physics the whole set of states \Math{\wS} usually has the special structure
of a set of functions \Math{\wS=\lSX} on some \emph{space} \Math{\X} with
values in some set of \emph{local states} \Math{\lS}. 
In dynamical systems with space nontrivial gauge structures arise naturally. 
The gauge structures are used in physical theories for description
of forces.
\par
We assume that the space is a finite set \Math{\X=\set{\x_1,\ldots, \x_\XN}}.
Its symmetries form the \emph{group of space symmetries} 
\Math{\sG=\set{\sg_1=\id,\ldots,\sg_\sGN}\leq\Perm{\X}}.
The case when \Math{\sG} is a \emph{proper} subgroup of \Math{\Perm{\X}}
implies that \Math{\X} possesses some additional structure. For example,
--- and this is sufficient for our purposes --- \Math{\X} may be an 
\emph{abstract} graph.
\par
The local states form a finite set \Math{\lS=\set{\ls_1,\ldots,\ls_\lSN}}
provided with the \emph{group of internal symmetries} 
\Math{\iG=\set{\ig_1=\id,\ldots, \ig_\iGN}\leq\Perm{\lS}}.
\par
There are different ways to combine the space \Math{\sG} 
and internal \Math{\iG} groups into the symmetry group \Math{\wG} of the whole
set of states \Math{\wS=\lSX}. The following equivalence class of 
\emph{split extensions}
is a natural generalization of constructions used in physical theories
\begin{equation}
\id\rightarrow\iGX\rightarrow\wG\rightarrow\sG\rightarrow\id.
\label{extentionEn}
\end{equation}
Here \Math{\iGX} is the group of \Math{\iG}-valued functions on the space \Math{\X}. 
Explicit formulas expressing group operations in \Math{\wG} from 
\eqref{extentionEn} in terms of operations in \Math{\sG} and \Math{\iG} 
are presented in \cite{KornyakJMS10En, KornyakNS10En} --- we do not need them
in this paper.
%___________________________________________________________________________
\subsection{Notational remarks}
In view of our further purposes, it is convenient for us to include zero
in the set of natural numbers, i.e., we shall use the definition: 
\Math{\Nat=\set{0,1,2,\ldots}}.
\par
If it is necessary to indicate explicitly whether an element \Math{\psi} 
of a Hilbert space \Math{\Hspace} is vector or covector%
\footnote{Any Hilbert space is canonically isomorphic to its dual as a space with inner
product.}, we shall use the notations \Math{\barket{\psi}} and \Math{\brabar{\psi}},
respectively. 
\par
For the \emph{standard inner product} in \Math{\adimH}-dimensional Hilbert space
we use the round brackets:
\begin{equation}
	\innerstandard{\phi}{\psi}\equiv\sum\limits_{i=1}^{\adimH}\cconj{\phi^i}\psi^i.
\label{innerstdEn}
\end{equation}
\par
For the \emph{invariant inner product} the angle brackets are used:
\begin{equation}
\inner{\phi}{\psi}\equiv\frac{\textstyle{1}}{\textstyle{\cabs{G}}}\sum\limits_{g\in{}G}
\!\innerstandard{U\!\vect{g}\phi}{U\vect{g}\psi},
\label{innerinvEn}
\end{equation}
where \Math{U} is a representation of a group \Math{G} in the space \Math{\Hspace}.

%_____________________________________________________________________
\section{Quantum Evolution of Dynamical System}
%_____________________________________________________________________
The most popular and intuitive method of quantization is
Feynman's path integral approach \cite{FeynmanEn}. 
This method is particularly well suited for dynamical systems with space structures.
According to the Feynman approach, the amplitude of quantum transition from initial
to final state is computed by summing up the amplitudes along all possible classical
trajectories connecting these states. The amplitude along a particular trajectory
is computed as the product of amplitudes of transitions between the nearest subsequent
states lying on the trajectory. Usually the amplitude is written as 
exponent of the action along the trajectory
\begin{equation}
A_{\mathrm{U}(1)}~=~A_0\exp\vect{iS}=A_0\exp\vect{i\int\limits_0^T{}Ldt}.
\label{amplclassEn}
\end{equation}
The function \Math{L}, depending on first order time derivatives of states,
is called \emph{Lagrangian}. In the discrete time the exponent of the integral
turns into the product 
\Math{\exp\vect{i\int{}Ldt}\rightarrow\e^{{iL_{0,1}}}\ldots\e^{{iL_{t-1,t}}}
	\ldots\e^{{iL_{T-1,T}}}} and the amplitude takes the form
\begin{equation}
	A_{\mathrm{U}(1)}~=~A_0
	\e^{{iL_{0,1}}}\ldots\e^{{iL_{t-1,t}}}
	\ldots\e^{{iL_{T-1,T}}}.\label{ampldiscrEn}
\end{equation}
It is natural to interpret the factors \Math{\Partransport_{t-1,t}=\e^{{iL_{t-1,t}}}}
of this product as \emph{connections} (\emph{parallel transports}) with
values in \emph{one}-dimensional unitary representation of the circle, i.e.,
commutative Lie group \Math{\iG=S^1\equiv\R/\Z}.
\par 
Assuming that the group \Math{\iG} is not necessarily \Math{S^1} and that its 
representation \Math{\Rep{\iG}} is not necessarily one-dimensional, we obtain a natural
ge\-ne\-ra\-li\-za\-tion of \eqref{ampldiscrEn}%
\footnote{In the non-commutative case we should observe the correct order of operators
 --- compatible in this particular context with the tradition 
to write matrices on the left of vectors.}
\begin{equation}
	A_{\Rep{\iG}}=\Rep{\alpha_{T,T-1}}\ldots\Rep{\alpha_{t,t-1}}
	\ldots\Rep{\alpha_{1,0}}A_0,\hspace*{10pt}\alpha_{t,t-1}\in\iG.\label{amplgenEn}
\end{equation}
Now the amplitude is a multicomponent vector suitable for description of particles having
internal degrees of freedom and moving in space. We shall assume that \Math{\iG} is
a finite group --- recall that linear representations of finite groups are unitary 
automatically. 
It is clear, that the standard quantization \eqref{amplclassEn} can be approximated 
via one-dimensional representations of finite cyclic groups.
\par
As is well known, Feynman's approach is equivalent to the traditional matrix formulation
of quantum mechanics. According to the matrix formulation, the evolution of a system
from the initial to final state is described by the evolution matrix \Math{U}:
\Math{\barket{\psi_{\tin}}\rightarrow\barket{\psi_{\tfin}}=U\barket{\psi_{\tin}}}.
The evolution matrix can be represented as the product of matrices corresponding to the
elementary time steps: 
\Mathh{U=U_{\tfin\leftarrow\tfin-1}\cdots{}U_{t\leftarrow{}t-1}\cdots{}U_{1\leftarrow0}.}
In fact, Feynman's quantization rules --- ``multiply subsequent events'' and
``sum up alternative histories'' --- is simply a rephrasing of the matrix
 multiplication rule. This is clear from the below illustration, where two steps of 
 evolution of a two-state quantum system (\emph{one-qubit register}) are presented
 in parallel in both Feynman's and matrix forms:
\begin{center}
\begin{tabular}[t]{ccc}
\includegraphics[height=0.24\textwidth]{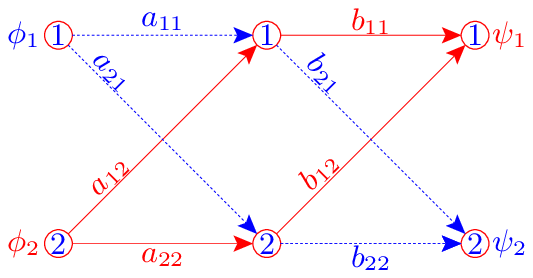}
&\raisebox{0.1\textwidth}{$\sim$}%{$\Huge\Longleftrightarrow$}
&
\includegraphics[height=0.24\textwidth]{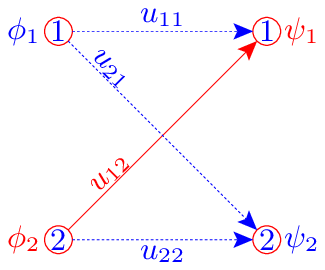}
\\
$\Updownarrow$&&$\Updownarrow$
\\
$
BA=
\begin{bmatrix}
{\color{blue}b_{11}a_{11}+b_{12}a_{21}}
&
{\color{red}b_{11}a_{12}+b_{12}a_{22}}
\\[5pt]
{\color{blue}b_{21}a_{11}+b_{22}a_{21}}
&
{\color{blue}b_{21}a_{12}+b_{22}a_{22}}
\end{bmatrix}
$~~~~~~~~~
& $\sim$~~ &
$
U=
\begin{bmatrix}
{\color{blue}u_{11}}&{\color{red}u_{12}}
\\[5pt]
{\color{blue}u_{21}}&{\color{blue}u_{22}}
\end{bmatrix}
$~~~~~~~~
\end{tabular}
\end{center}
According to the Feynman rules, the transition, say, between the states
\Math{\phi_2} and \Math{\psi_1} is the sum \Math{b_{11}a_{12}+b_{12}a_{22}}
over two paths. But this is just the element 
\Math{u_{12}} of the matrix product \Math{U=BA}.
The general case of many time steps and many states can easily be deduced from 
this elementary example by induction.
\par
The above reasoning works also in the case of non-commutative gauge
connection as in \eqref{amplgenEn}. 
We should only treat the evolution matrices \Math{A, B} and \Math{U}
as block matrices with non-commutative entries which are matrices from the
representation \Math{\Rep{\iG}}. For the sake of uniformity of consideration
we can ignore this block structure and interpret the matrices as ordinary
matrices of larger dimension from representations of the whole symmetry group \Math{\wG}
constructed in accordance with \eqref{extentionEn}.
\par
In quantum mechanics, evolution matrices \Math{U} are unitary operators acting
in Hilbert spaces of \emph{state vectors} 
(called also ``\emph{wave functions}'', ``\emph{amplitudes}'' etc.).
\emph{Quantum mechanical particles} are associated with unitary representations 
of certain groups. These representations are called ``\emph{singlets}'', 
``\emph{doublets}'', and so on, in accordance with their dimensions.
Multidimensional representations describe the \emph{spin}.
A \emph{quantum mechanical experiment} is reduced to comparison of the system
state vector \Math{\psi} with some sample state vector \Math{\phi} 
provided by a ``\emph{measuring apparatus}''.
According to the Born rule, the probability to observe the coincidence of the states
is equal to \Math{{\cabs{\inner{\phi}{\psi}}^2}} 
(assuming the normalization \Math{\inner{\phi}{\phi}=\inner{\psi}{\psi}=1}).
To make all these quantum %mechanical 
concepts constructive, we suppose 
 that the evolution operators are elements of representations of finite groups.
%__________________________________________________________________________________________
\section{Quantum Description of Finite Systems}
%__________________________________________________________________________________________
\subsection{Permutations and linear representations}
\subsubsection{Group actions.}
All transitive actions of a finite group \Math{\wG=\set{\wg_1,\ldots,\wg_{\wGN}}}
on finite sets \Math{\Omega=\set{\omega_1,\ldots,\omega_n}}
can easily be described \cite{HallEn}. 
Any such set is in one-to-one correspondence with the \emph{right} (or \emph{left}) 
\emph{cosets} of some subgroup \Math{H\leq\wG}, i.e., 
\Math{\Omega\cong{}H\backslash\wG} (or \Math{\Omega\cong{}\wG/H}).
The set \Math{\Omega} is called the \emph{homogeneous space} of the group \Math{\wG} 
(or \Math{\wG}-\emph{space}). 
Action of \Math{\wG} on \Math{\Omega} is \emph{faithful}, if the subgroup \Math{H}
does not contain normal subgroups of \Math{\wG}. We can write action in the form of
permutations

\Mathh{
\pi(g)=\dbinom{\omega_i}{\omega_ig}\sim\dbinom{Ha}{Hag},
\hspace*{20pt}g,a\in{}\wG,~~~i=1,\ldots,n.
}
\par
Maximal transitive set \Math{\Omega} is the set of all elements of the group \Math{\wG}
itself, i.e., the set of cosets of the trivial subgroup \Math{H=\set{\id}}. 
The corresponding action is called \emph{regular} and can be represented by the permutations
\begin{equation}
	\Pi(g)=\dbinom{\wg_i}{\wg_ig},~~~~i=1,\ldots,\wGN.
	\label{regpermEn}
\end{equation}
\par
To introduce a ``quantitative'' (``statistical'') description, let us assign
to the elements of the set \Math{\Omega} numerical ``weights'' from some suitable
\emph{number system} \Math{\aNumbers} containing at least \emph{zero} and \emph{unity}.
This allows rewriting permutations in the matrix form 
\begin{equation}
\pi(g)\rightarrow\rho(g)=
\Mone{\rho(g)_{ij}},\text{~~ where~~} \rho(g)_{ij}=\delta_{\omega_ig,\omega_j};~~ i,j=1,\ldots,n;
\label{permrepEn}
\end{equation}
\Mathh{\delta_{\alpha,\beta}\equiv
\begin{cases}
1, & \text{if~~} \alpha=\beta,\\
0, & \text{if~~} \alpha\neq\beta,
\end{cases}
\text{~~for~~} \alpha,\beta\in\Omega.
} 
The function \Math{\rho}, defined by \eqref{permrepEn},
is called the \emph{permutation representation}.
\par
The \emph{cycle type} of a permutation is the array of multiplicities
of lengths of cycles in the decomposition of the permutation into disjoint
cycles. The cycle type is usually denoted by
\Math{1^{k_1}2^{k_2}\cdots{}n^{k_n},}
where \Math{k_i} is the number of cycles of the length \Math{i} in the permutation; 
\Math{0\leq{}k_i\leq{}n;~~k_1+2k_2+\cdots+nk_n=n.}
The \emph{characteristic polynomial} of permutation matrix \eqref{permrepEn} can 
be written immediately from the cycle type of the corresponding permutation \Math{\pi(g)}:
\begin{equation}
\chi_{\rho(g)}\vect{\lambda}=\det\vect{\rho(g)-\lambda\idmat}
=\vect{\lambda-1}^{k_1}\vect{\lambda^2-1}^{k_2}\cdots\vect{\lambda^n-1}^{k_n}.
\label{charpolEn}	
\end{equation}
The matrix form of permutations \eqref{regpermEn} representing the
 \emph{regular} action
\begin{equation}
	\Pi(g)\rightarrow{}\regrep(g)=
\Mone{\regrep(g)_{ij}},~~ \regrep(g)_{ij}=\delta_{e_ig,e_j},~~ i,j=1,\ldots,\wGN
	\label{regrepEn}
\end{equation}
is called the \emph{regular representation} --- this is a special case of more general 
permutation representation \eqref{permrepEn}.
\par
For the sake of freedom of algebraic manipulations, one assumes usually that
\Math{\aNumbers} is an algebraically closed field --- for example, 
the field of complex numbers \Math{\C}.
%More ``economical'' number system used in the theory of finite groups is extension 
%of the field of rationals by a primitive root of unity: \Math{\aNumbers=\Q\vect{\runi{}}}.
If \Math{\aNumbers} is a field, then the set \Math{\Omega} 
can be treated as a basis of linear vector space	
\Math{\Hspace=\mathrm{Span}\vect{\omega_1,\cdots,\omega_n}}.
%____________________________________________________________________________
\subsubsection{Number systems}
As is clear from \eqref{charpolEn}, all \emph{eigenvalues} of permutation matrices are roots
of unity. This --- in combination with the fact that all irreducible representations
of finite groups are subrepresentations of regular representations \eqref{regrepEn} 
--- means that all numbers sufficient for our purposes can be constructed from
the \emph{natural numbers} \Math{\N=\set{0,1,\ldots}} and primitive \emph{root of unity} 
\Math{\runi{}} of a certain degree \Math{n}.
The term \emph{primitive} means that \Math{\runi{}^n=1} and period of 
\Math{\runi{}} is equal exactly to \Math{n}. 
As the degree \Math{n} one can always take the \emph{exponent} of the group \Math{\wG}
--- the least common multiple of orders of the group elements. But in many cases
some proper divisor of the exponent is enough. Any root of unity can be expressed via
the primitive root as
 \Math{\runi{}^k,~k\in\set{0,1,\ldots,n-1}}.
For intuitive perception one could bear in mind the 
symbolics \Math{\runi{}=\e^{2\pi{}i/n}} for the primitive root, 
but we will never use this representation.
The following \emph{algebraic} definitions are sufficient for all our purposes 
\begin{enumerate}
	\item \emph{Multiplication}: \Math{\runi{}^k\times\runi{}^m=\runi{}^{k+m\mod{}n}},
	\item \emph{Complex conjugation}: \Math{\cconj{\runi{}^k}=\runi{}^{n-k}}.
\end{enumerate}
Thus we will use as a \emph{number system} the set of polynomials in \Math{\runi{}} with
\emph{natural} coefficients: \Math{\aNumbers=\N\ordset{\runi{}}}.
\par
If \Math{n=1}, then
\Math{\aNumbers} is the \emph{semi-ring of natural numbers} \Math{\N}.
This number system corresponds to the case of trivial group \Math{\wG}.
\par 
If \Math{n\geq2}, then the \emph{negative numbers} can be introduced via the definition 
\Mathh{\vect{-1}=
\begin{cases}
\runi{}^{n/2}, & \text{if~}n\text{~is~even},\\
\runi{}+\cdots+\runi{}^{n-1}, & \text{if~}n\text{~is~odd.}
\end{cases}
}
In particular, at \Math{n=2} the \emph{ring of integers} arises: \Math{\aNumbers=\Z}.
This numerical system corresponds to the case \Math{\wG=\CyclG{2}} 
\Math{\vect{\text{or~}\wG=\CyclG{2}\times\cdots\times\CyclG{2}}}. 
\par
If \Math{n\geq3}, then the set 
\Math{\aNumbers=\N\ordset{\runi{}}} is a \emph{commutative ring} embeddable into 
the field of complex numbers \Math{\C}.
\par
The ring \Math{\N\ordset{\runi{}}} is sufficient for all computations with the
finite quantum models. For descriptive simplicity of the linear algebra
--- in order to we could talk freely about linear spaces ---
we will use also the quotient field%
\footnote{From a computational point of view, working over rings --- though it slightly 
complicates algorithms --- is more efficient than computations over corresponding
quotient fields. Modern computer algebra systems use mainly the ring computations
in the linear algebra algorithms.}
 of this ring. Sometimes square roots of natural numbers
may arise as intermediate technical symbols (in fact, square roots of naturals
can always be expressed in terms of roots of unity, but we do not need this 
possibility).
Neither roots of unity nor other irrationalities appear in the final expressions
having the status of ``observables''.
%__________________________________________________________________
\subsubsection{Unitary representations}
\label{unirep}
Unitary operators play a key role in quantum mechanics.
Any linear representation of a finite group is equivalent to unitary,
since one can always construct invariant inner product from an arbitrary one by
``averaging over the group''. For example, invariant product 
\eqref{innerinvEn} is constructed by averaging the standard product defined in \eqref{innerstdEn}.
\par
Let us describe briefly the main facts \cite{HallEn} about irreducible representation of
finite groups, illustrating them with the help of the smallest
non-commutative group --- the group of permutations of three elements \Math{\SymG{3}}.
It is also isomorphic to the symmetry group of a triangle, i.e., to the dihedral group 
\Math{\DihG{6}\cong\SymG{3}}. The group consists of six elements
having the following representation by permutations
\begin{equation}
\wg_1=\vect{},~\wg_2=\vect{2,3},~\wg_3=\vect{1,3},~\wg_4=\vect{1,2},
	~\wg_5=\vect{1,2,3},~\wg_6=\vect{1,3,2}.
\label{S3elemsEn}	
\end{equation}
\par
An important transformation of group elements
--- an analog of change of coordinates in physics --- is the conjugation:
\Math{a^{-1}ga\rightarrow{}g',} \Math{g, g'\in\wG,} \Math{a\in\Aut{\wG}}.
Conjugation by an element of the group itself, i.e., if \Math{a\in\wG},
is called an \emph{inner automorphism}.
The equivalence classes with respect to the inner automorphisms are called
\emph{conjugacy classes}. Decomposition of a group into conjugacy classes,
symbolically written as
\Mathh{\wG=\class{1}+\class{2}+\cdots+\class{\classN},}
plays an important role in the study of its representations.
\par
\emph{Example.} The group \Math{\SymG{3}} decomposes into three congugacy classes
\begin{equation}
	\class{1}=\set{\vect{}},~~\class{2}=\set{\vect{2,3},~\vect{1,3},~\vect{1,2}},
	~~\class{3}=\set{\vect{1,2,3},~\vect{1,3,2}}.
	\label{S3classesEn}
\end{equation}
\par
The group multiplication induces \emph{multiplication} of the classes.
The product of two classes \Math{\class{i}} and \Math{\class{j}} is 
the \emph{multiset} of all possible products 
\Math{ab,~a\in\class{i},~b\in\class{j}} decomposed into classes.
This multiplication is obviously commutative, since \Math{ab} and \Math{ba} belong to 
the same class: \Math{ab\sim{}a^{-1}\vect{ab}a=ba}. Thus, the multiplication
table for classes is given by
\begin{equation}
	\class{i}\class{j}= \class{j}\class{i} = \sum\limits_{k=1}^{\classN}c_{ijk}\class{k}.
\label{classtabEn}	
\end{equation}
The \emph{natural integers} \Math{c_{ijk}} 
--- multiplicities of classes in the multisets --- are called
\emph{class coefficients}.
\par
\emph{Example.} The group \Math{\SymG{3}} has the following multiplication table for classes
\Mathh{\class{1}\class{j}=\class{j},~~\class{2}^2=3\class{1}+3\class{3},
~~\class{2}\class{3}=2\class{2},~~\class{3}^2=2\class{1}+\class{3}.}
\par
Some of the main properties of linear representations of finite groups are listed below:
\begin{enumerate}
	\item 
Any linear representation is (equivalent to) unitary.
	\item 
Every possible irreducible representation is contained in the regular 
re\-pre\-sen\-ta\-tion.
More specifically,	there exists (unitary) matrix \Math{\transmatr} transforming
si\-mul\-ta\-neously all matrices
\eqref{regrepEn} to the form
\begin{equation}
	\transmatr^{-1}\regrep(g)\transmatr=
	\bmat
			\repirr_1(g) &&&
	\\[5pt]
	&
	\hspace*{-27pt}d_2\left\{
	\begin{matrix}
	\repirr_2(g)&&\\
	&\hspace*{-10pt}\ddots&\\
	&&\hspace*{-7pt}\repirr_2(g)
	\end{matrix}
	\right. 
	&&
	\\
	&&\hspace*{-10pt}\ddots&\\
	&&& 
	\hspace*{-25pt}d_{\classN}\left\{
	\begin{matrix}
	\repirr_{\classN}(g)&&\\
	&\hspace*{-10pt}\ddots&\\
	&&\hspace*{-7pt}\repirr_{\classN}(g)
	\end{matrix}
	\right. 
	\emat
\label{regrepdecompEn}	
\end{equation}
and any irreducible representation is one of the elements of the set
\Math{\set{\repirr_1,\ldots,\repirr_{\classN}}}.
The number of non-equivalent irreducible representation \Math{\classN}
is equal to the number of conjugacy classes \Math{\class{j}} 
in the group \Math{\wG}. The number \Math{d_j} is the dimension of 
the irreducible component \Math{\repirr_j} and simultaneously the multiplicity
of its occurrence in the regular representation.
It is clear from \eqref{regrepdecompEn}, that
for the dimensions of irreducible representations the following relation holds:
%the sum of squares of dimensions 
%of irreducible representations is equal to the order of the group: 
\Math{d^2_1+d^2_2+\cdots+d^2_{\classN}=\cabs{\wG}=\wGN.}
It can be proved also that the dimensions of irreducible representations divide
the group order:
\Math{d_j\mid\wGN.}
	\item 
Any irreducible representation \Math{\repirr_j} is determined uniquely
(up to isomorphism) by its \emph{character} \Math{\chi_j}. This is a function
on the group defined as the trace of the representation matrix: 
\Math{\chi_j\vect{g}=\mathrm{Tr}\repirr_j\vect{g}}.
In fact, the character is a function on the conjugacy classes since
\Math{\chi_j\vect{g}=\chi_j\vect{a^{-1}ga}}.
Such functions are called \emph{central}, and any central function is a 
linear combination of characters. The value of character \Math{\chi_j} on the class 
\Math{\class{1}=\set{\id}} is equal to the dimension \Math{d_j}.
	\item
A compact form of recording all irreducible representations is the 
\emph{character table}.
The columns of this table are numbered by the congugacy classes,
while its rows contain values of characters of non-equivalent representation:
\begin{center}
\begin{tabular}{c|cccc}
&\Math{\class{1}}&\Math{\class{2}}&\Math{\cdots}&\Math{\class{\classN}}\\\hline
\Math{\chi_1}&1&1&\Math{\cdots}&1\\
\Math{\chi_2}&\Math{\chi_2\vect{\class{1}}=d_2}&\Math{\chi_2\vect{\class{2}}}&
\Math{\cdots}&\Math{\chi_2\vect{\class{\classN}}}\\
\Math{\vdots}&\Math{\vdots}&\Math{\vdots}&&\Math{\vdots}\\
\Math{\chi_{\classN}}&\Math{\chi_{\classN}\vect{\class{1}}=d_{\classN}}&
\Math{\chi_{\classN}\vect{\class{2}}}&\Math{\cdots}&\Math{\chi_{\classN}\vect{\class{\classN}}}
\end{tabular}.
\end{center}
According to the standard convention, the 1st column corresponds to the class of the group 
identity, and the 1st row contains the \emph{trivial} representation.
\end{enumerate}
%_________________________________________________________________
\subsection{Embedding of quantum system into classical}
In the most general formulation, quantum mechanics assumes that every physical
system corresponds to a Hilbert space \Math{\Hspace} non-zero vectors of which,
 \Math{\psi\in\Hspace}, represent all possible states of the system.
It is assumed also that vectors \Math{\psi} and \Math{\psi'}
describe identical states if they are proportional through a complex factor: 
\Math{\psi'=\lambda\psi,~\lambda\in\C}.
Evolution of the system from any initial state \Math{\psi_0} into 
the corresponding final state \Math{\psi_T} is described by an \emph{unitary} 
operator \Math{U}: \Math{\barket{\psi_T}=U\barket{\psi_0}}.
The unitarity means that \Math{U} belongs to the automorphism group 
of the Hilbert space: \Math{U\in\Aut{\Hspace}}.
One may regard \Math{\Aut{\Hspace}} as a faithful unitary representation of
respective abstract group \Math{\wG}.
In the continuous time the dynamics can be expressed by 
the Schr\"{o}dinger equation 
\Mathh{i\frac{\mathrm{d}}{\mathrm{d}t}\barket{\psi}=H\barket{\psi}}
in terms of the local operator \Math{H}
called the \emph{Hamiltonian} or \emph{energy operator}.
If the \emph{Hermitian} operator \Math{H} is independent of time,
then it is connected with the \emph{unitary} evolution operator \Math{U} by the
simple relation \Math{U=\e^{-iHT}.} 
\par
A finite quantum system is formulated in exactly the same way.
The only difference is that now the group \Math{\wG} is assumed to be a finite group
of order \Math{\wGN} having unitary representation \Math{\mathrm{U}}
in \Math{\adimH}-dimensional Hilbert space \Math{\Hspace_{\adimH}}.
All possible evolution operators form the finite set
\Math{\set{U_1,\ldots,U_{\wGN}}} of unitary matrices from \Math{\mathrm{U}}.
Since the matrices \Math{U_j} are non-singular, one can always introduce
Hamiltonians by the formula \Math{H_j=i\ln{}U_j}, but there is no need to do so.
Note that many applications of quantum mechanics --- 
e.g., \emph{quantum computing}
or \Math{S}-\emph{matrix} approach --- mostly avoid use of Hamiltonians.
\par
Finite groups --- if they are ``sufficiently non-commutative''
--- can be often generated by a small number of elements.
For example, all simple and all symmetric groups are generated by two elements.
The algorithm restoring the whole group from \Math{n_{g}} generators
is very simple. It is reduced to \Math{n_{g}\vect{\wGN-n_{g}-1}} group multiplications.
So the finite quantum models are well suited for study by the
computer algebra methods.
%\par
%To avoid inessential technical complications, we may assume that the
%representation \Math{\mathrm{U}} is irreducible.
\def\PAR{\par}
\PAR
It follows from decomposition \eqref{regrepdecompEn} 
that any \Math{\adimH}-dimensional representation \Math{\mathrm{U}}
can be extended to an \Math{\wSN}-dimensional representation
\Math{\mathrm{\widetilde{U}}} in a Hilbert space \Math{\Hspace_{\wSN}},
in such a way that the representation \Math{\mathrm{\widetilde{U}}} 
corresponds to the \emph{permutation action} of the group \Math{\wG} on 
some \Math{\wSN}-element set of states \Math{\wS=\set{\ws_1,\ldots,\ws_{\wSN}}}.
It is clear that \Math{\wSN\geq\adimH}. 
\PAR
The case when \Math{\wSN} is strictly greater than \Math{\adimH}
is most interesting.
Clearly, the additional ``hidden parameters''
 --- appearing in this case due to increase of the number of states (dimension of space) ---
in no way can affect the data relating to the %initial 
space \Math{\Hspace_{\adimH}}
since both \Math{\Hspace_{\adimH}} and its complement in \Math{\Hspace_{\wSN}} 
are invariant subspaces of the extended space \Math{\Hspace_{\wSN}}.
%\PAR
Thus \emph{any quantum problem} in \Math{\adimH}-dimensional Hilbert space
can be reduced to permutations of \Math{\wSN} things.
\PAR
From the algorithmic point of view, manipulations with permutations are much more 
efficient than the linear algebra operations with matrices. On the other hand,
degrees of permutations \Math{\wSN} might be much larger than dimensions of matrices
 \Math{\adimH}. However, the possibility to \emph{reduce quantum dynamics to permutations}
 in principle 
 is much more important from the conceptual point of view than the algorithmic issues.
\paragraph
{Example.}
\textbf{The group \Math{\SymG{3}}} has the following character table
\begin{equation}
	\text{\begin{tabular}{c|crr}
	&\Math{\class{1}}&\Math{\class{2}}&\Math{\class{3}}\\\hline
	\Math{\chi_1}&1&1&1\\
	\Math{\chi_2}&1&-1&1\\
	\Math{\chi_3}&2&0&-1
	\end{tabular}\enspace.}\label{S3tabEn}
\end{equation}
The contents of the classes \Math{\class{1}, \class{2}, \class{3}} are 
presented in \eqref{S3classesEn}.
As the representation \Math{\mathrm{U}} describing quantum evolution,
let us take the two-dimensional faithful representation corresponding to 
the character \Math{\chi_3}.
The matrices (evolution operators) of this representation
--- ordered in accordance with 
%the order of the group elements presented in 
\eqref{S3elemsEn} 
--- are the following
\begin{align}
		U_1=\Mtwo{1}{0}{0}{1},~U_2=\Mtwo{0}{\runi{}^2}{\runi{}}{0},
		~U_3=\Mtwo{0}{\runi{}}{\runi{}^2}{0},\nonumber\\[-6pt]
		\label{S3umatsEn}\\[-6pt]
		~U_4=\Mtwo{0}{1}{1}{0},
		~U_5=\Mtwo{\runi{}^2}{0}{0}{\runi{}},~U_6=\Mtwo{\runi{}}{0}{0}{\runi{}^2},\nonumber
	\end{align}
where \Math{\runi{}} is a primitive third root of unity.%
\footnote{Note the peculiarity of representation \eqref{S3umatsEn}
--- its matrices are very similar to matrices of permutations: there is exactly
one non-zero entry in each column and in each row. But in contrast to permutation
matrices in which any non-zero entry is \emph{unity}, 
non-zeros in \eqref{S3umatsEn} are \emph{roots of unity}.
This is because \Math{\SymG{3}} is one of the so-called
\emph{\textbf{monomial groups}} for which all irreducible representations can be
 constructed as induced from one-dimensional representations of their subgroups
(the relevant to \eqref{S3umatsEn} subgroup is \Math{\CyclG{3}\leq\SymG{3}}). 
Note also, that most groups, at least of small orders,
are just monomial. For example, the total number of all non-isomorphic groups
of order \Math{<384} is equal to 67424, but only 249 of them are \emph{\textbf{non-monomial}}.
The minimal non-monomial group is the 24-element group \Math{\SL{2}{3}} of \Math{2\times2} 
matrices in the characteristic 3 with unit determinants.}
\paragraph{Quantum and permutation matrices for \Math{\SymG{3}}.}
Any permutation re\-pre\-sen\-ta\-tion contains one-dimensional invariant subspace with 
the basis vector \Math{\Vthree{1}{\vdots}{1}}.
So there is only one way to extend 
\eqref{S3umatsEn} to the permutations of three elements --- we must add the trivial
representation corresponding to the character \Math{\chi_1} in \eqref{S3tabEn}
---
and come to the 3-dimensional representation \Math{\mathrm{\widetilde{U}}} with
the matrices
\begin{equation}
	\widetilde{U}_j=\Mtwo{1}{0}{0}{U_j},~~j =1,\ldots,6.
	\label{S3permqEn}
\end{equation}
These matrices are permutation matrices in the basis in which the permutation
representation is decomposed into invariant components. We shall call such bases
\emph{quantum}.
In the \emph{permutation} basis matrices \eqref{S3permqEn} are, respectively, the following
\begin{align}
		P_1=\Mthree{1}{\cdot}{\cdot}{\cdot}{1}{\cdot}{\cdot}{\cdot}{1},
		~P_2=\Mthree{1}{\cdot}{\cdot}{\cdot}{\cdot}{1}{\cdot}{1}{\cdot},
		~P_3=\Mthree{\cdot}{\cdot}{1}{\cdot}{1}{\cdot}{1}{\cdot}{\cdot},\nonumber\\[-6pt]
		\label{S3pmatsEn}\\[-6pt]
		~P_4=\Mthree{\cdot}{1}{\cdot}{1}{\cdot}{\cdot}{\cdot}{\cdot}{1},
		~P_5=\Mthree{\cdot}{1}{\cdot}{\cdot}{\cdot}{1}{1}{\cdot}{\cdot},
		~P_6=\Mthree{\cdot}{\cdot}{1}{1}{\cdot}{\cdot}{\cdot}{1}{\cdot}.\nonumber
	\end{align}
The most general unitary matrix of transition from the permutation basis 
to the quantum --- we define this transformation matrix via the relation 
\Math{\widetilde{U}_j=\transmatr^{-1}P_j\transmatr}
--- takes the form	
\Mathh{
\transmatr=\frac{1}{\sqrt{3}}
\Mthree{a}{b~~}{b\runisymb^2}
 {a}{b\runisymb^2}{b~~}
 {a}{b\runisymb}{b\runisymb},
}
where \Math{a} and \Math{b} are arbitrary elements of the set 
\Math{\set{1,\runisymb,\runisymb^2}}.
The concrete choice of these elements is absolutely inessential, since they dissapear 
in the inner products included in the expressions for observables.
So finally we take the following form for the transformation matrix
\begin{equation}
\transmatr=\frac{1}{\sqrt{3}}
	\Mthree{1}{1}{\runisymb^2}
	 {1}{\runisymb^2}{1}
	 {1}{\runisymb}{\runisymb},~~~~
\transmatr^{-1}=\frac{1}{\sqrt{3}}
	\Mthree{1}{1}{1}
	 {1}{\runisymb}{\runisymb^2}
	 {\runisymb}{1}{\runisymb^2}.
\label{transS3En}	 
\end{equation}
As we shall see later, all information about the ``quantum behavior'' of the
group \Math{\SymG{3}} acting on the set of 3 elements is encoded, in fact, 
in this matrix.
\subsubsection{On simulation of quantum computing via finite groups}
Designing a quantum algorithm is reduced to constructing unitary operator 
realizing the algorithm from a given set of standard operators (\emph{gates}).
There are universal sets of gates.
The \emph{universality} means that any unitary operator can be approximated 
by combination of operators from this set, i.e., the gates are \emph{generators}
of finitely generated group which is dense in the set of all
unitary operators acting on corresponding quantum register.
Let us consider, e.g., the following set of operators:\\
(a) \emph{the Hadamard gate} \Math{H=\frac{1}{\sqrt{2}}\Mtwo{1}{1}{1}{-1}},~
(b) ``\emph{the phase shifter (\emph{or} phase rotating) gate}''	
\Math{R\vect{\theta}=\Mtwo{1}{0}{0}{\e^{2\pi{}i\theta}}}
and (c) ``\emph{the controlled-NOT gate}'' 
\Math{\text{CNOT}=\bmat1&0&0&0\\0&1&0&0\\0&0&0&1\\0&0&1&0\emat}.
At certain values of parameter \Math{\theta} these operators generate finite groups.
For example, at \Math{\theta=1/4}, on two-qubit register, the generated group \Math{G}
has the size \Math{36864}. The computer algebra system \textbf{GAP} 
\cite{gapEn} gives the following structure for this group: 
\Mathh{G\cong\vect{\vect{\vect{\vect{\CyclG{8}\times\CyclG{2}}\rtimes\CyclG{2}}
\rtimes\CyclG{3}}\rtimes\CyclG{2}}
\times\vect{\vect{\SL{2}{3}\rtimes\CyclG{4}}\rtimes\CyclG{2}}.}
\par
If the value of \Math{\theta} is in general position, the operators (a), (b) and (c)
form universal set of gates and, hence, generate infinite groups.
But these groups, in a certain sense, are close to finite ---
they are \emph{residually finite}.
\par
Recall that a group \Math{G} is called \emph{residually finite} \cite{MagnusEn}
if for each its element \Math{g\neq\id}, there is a homomorphism 
\Math{\phi: G\rightarrow{}H} onto a finite group \Math{H}, such that 
\Math{\phi\vect{g}\neq\id}. 
This means that any relations between the elements of \Math{G} 
can be simulated by relations between the elements of some finite group. 
According to Mal'cev's theorem: \emph{any finitely generated group of matrices
over a field is residually finite}.
\par 
Thus, any universal set of gates generates residually finite groups.
This makes it possible to simulate quantum computations by finite models
in analogy to the widely used in physics trick
 when to solve a problem, an infinite space
is replaced by a torus of the size sufficient to embrace the data of the problem.

\subsection{Connection of mathematical description with observation.\\ The Born rule.}
%\subsection{Правило Борна. Связь математического описания с наблюдением}
There are some subtleties in transition from mathematical description of 
systems with symmetries to observable ``objects'' or ``quantities''.
A detailed discussion of the subject can be found 
in the paper \cite{WeylEn}, and in the book \cite{ShafarevichEn} pp. 160 et seq.
In short, the point is that for registration and identification of elements of a system,
arbitrarily chosen marks are used.
Only such relations and statements have objective meaning as are not dependent on any change
in the choice of the labels, since such a change is nothing more than a renaming.
In systems with symmetries ``objects'' forming ``homogeneous'' set
(more formally, lying on the same group orbit) have different labels, but they are
indistinguishable in an absolute sense.
It is possible to fix such objects only with respect to some additional system 
which appears as ``\emph{coordinate system}'', or ``\emph{observer}'', or 
``\emph{measuring device}''.
For example, no absolute objective meaning can be attached to points of space,
denoted (marked) as vectors \Math{\mathbf{a}} and \Math{\mathbf{b}}.
But the relation between the points denoted as \Math{\mathbf{b}-\mathbf{a}}
(or in more general group notation \Math{\mathbf{a^{-1}b}})
is meaningful.
This is an example of the typical situation where observable objects or relations
are group invariants depending on \emph{pairs of elements}.
One of the elements of such a pair is related to observed system and another is related to observer.
\par
In quantum mechanics, the link between mathematical description and experiment is provided 
by the \emph{Born rule} \cite{LandsmanEn}, stating that the \emph{probability}
to observe a quantum system being in the state \Math{\psi} by apparatus tuned to the state
\Math{\phi} is expressed by the number
\begin{equation}
\ProbBorn{\phi}{\psi} = \frac{\textstyle{\cabs{\inner{\phi}{\psi}}^2}}
{\textstyle{\inner{\phi}{\phi}\inner{\psi}{\psi}}}.
\label{BornEn}	
\end{equation}
This expression can be rewritten in a form including the pair ``system--apparatus''
in more symmetric way
\Mathh{\ProbBorn{\phi}{\psi} = 
\frac{\textstyle{\cabs{\inner{\phi}{\psi}}^2}}
{\textstyle{\cabs{\inner{\phi}{\psi}}^2+\Grassnorm{\phi\wedge\psi}^2}}.}
Here \Math{\phi\wedge\psi} is exterior (Grassmann) product of the vectors 
\Math{\phi} and \Math{\psi}, which is the
%\Math{\frac{{\adimH(\adimH-1})}{{2}}}
\Math{\adimH(\adimH-1)/2}-dimensional vector with the components in the unitary basis
\Math{\vect{\phi\wedge\psi}^{ij}=\phi^i\psi^j-\phi^j\psi^i}
and with the square of norm
\Mathh{\Grassnorm{\phi\wedge\psi}^2=\sum\limits_{i=1}^{\adimH-1}
\sum\limits_{j=i}^{\adimH}\cabs{{\phi^{i}}\psi^{j}-{\phi^{j}}\psi^{i}}^2.}
In quantum mechanics, it is usually assumed that the state vectors 
are normed, i.e.,
 \Math{\inner{\phi}{\phi}=\inner{\psi}{\psi}=1}, and the Born rule is written 
in the form 
 \Math{\ProbBorn{\phi}{\psi} = \cabs{\inner{\phi}{\psi}}^2}. This simplifies
computations.
As an illustration, we can easily check that the function 
\Math{\ProbBorn{\phi}{\psi}} satisfies the basic property of probability
 --- 
the sum of probabilities of all possible results of observations of the state 
 \Math{\psi} is equal to unity. 
Namely, for any orthonormal basis 
 \Math{\set{\onbeq_1,\ldots,\onbeq_\adimH}} in a Hilbert space
\Math{\Hspace} we have:
\Mathh{\sum\limits_{i=1}^\adimH{}\set{\ProbBorn{\onbeq_i}{\psi}=
\cabs{\inner{\onbeq_i}{\psi}}^2=
\inner{\psi}{\onbeq_i}\inner{\onbeq_i}{\psi}}=
\left\langle\psi\underbrace{\cabs{\sum\limits_{i=1}^\adimH{}
\barket{\onbeq_i}\brabar{\onbeq_i}}}_{\idmat}\emph{}
\psi\right\rangle=\inner{\psi}{\psi}\equiv1.}
However, we --- trying to stick to as simple as possible number system
--- will not use the normalization of vectors.
\par
There are many philosophical speculations concerning the concept of probability and its
interpretation. However, what is really used in practice is the 
\emph{frequency interpretation}: the probability is the ratio of the number of favorable
cases to the total number of cases.
In the case of finite sets there are no complications at all: the probability
is the rational number, the ratio of the number of singled out elements of a set
to the total number of elements of the set.
\par
It can be shown that if data about states of a system and apparatus
are represented in the permutation basis by \emph{natural numbers}, then formula
 \eqref{BornEn} gives \emph{rational numbers} in the invariant subspaces of
 the permutation representation also, in spite of 
 possible presence of cyclotomic numbers
 and irrationalities in the intermediate computations.
\par
Let us consider permutation action of the group
\Math{\wG=\set{\wg_1,\ldots,\wg_{\wGN}}} on the set of states
\Math{\wS=\set{\ws_1,\ldots,\ws_{\wSN}}}.
We will describe the states of the system and apparatus in the permutation
representation by the vectors 
\begin{equation}
	\barket{n} = \Vthree{n_1}{\vdots}{n_{\wSN}} \text{~and~} 
	\barket{m} = \Vthree{m_1}{\vdots}{m_{\wSN}},
	\label{natamplEn}
\end{equation} 
respectively.
It is natural to assume that \Math{n_i} and \Math{m_i} are natural numbers,
interpreting them as the ``multiplicities of occurrences'' of the element
\Math{\ws_i} in the system and apparatus states, respectively.
Of course, due to the symmetry these numbers are not observable.
Only their \emph{invariant combinations} are observable.
Since the standard inner product defined in \eqref{innerstdEn} 
is invariant for the permutation representation, in accordance with the Born rule
we have
\begin{equation}
	\ProbBorn{m}{n}=\frac{\vect{\sum_i{m_i}n_i}^2}{\sum_i{m_i}^2\sum_i{n_i}^2}.
\label{probpEn}	
\end{equation}
It is clear that for non-vanishing vectors
 \Math{n} and \Math{m} with the natural components expression
\eqref{probpEn} is a rational number strictly greater than zero.
This means, in particular, that it is impossible to observe destructive quantum
interference here.
\par
However, the destructive interference of the vectors with natural components
can be observed in the proper invariant subspaces of the permutation representation.
Moreover, non-zero probabilities --- observable in the invariant subspaces --- 
are rational numbers.
Let us illustrate this by an example.
\paragraph{Illustration: Group \Math{\SymG{3}} acting on three elements.}
The state vectors in the permutation basis are 
\Math{\barket{n} = \Vthree{n_1}{n_2}{n_3}} and \Math{\barket{m} = \Vthree{m_1}{m_2}{m_3}}.
With the help of unitary transformation matrix
 \eqref{transS3En} 
we can transform the system state vector
\Math{n} 
from the permutation basis to the quantum basis
\Mathh{
\barket{\widetilde{\psi}}=\transmatr^{-1}\barket{n}
=\frac{1}{\sqrt{3}}
\Mthree{1}{1}{1}
 {1}{\runisymb}{\runisymb^2}
 {\runisymb}{1}{\runisymb^2}\Vthree{n_1}{n_2}{n_3}
=\frac{1}{\sqrt{3}}\Vthree{n_1+n_2+n_3}
{n_1+n_2\runisymb+n_3\runisymb^2}{n_1\runisymb+n_2+n_3\runisymb^2}.
}
The apparatus vector \Math{m} is transformed in the same way:
\Mathh{
\barket{\widetilde{\phi}}=\transmatr^{-1}\barket{m}
=\frac{1}{\sqrt{3}}\Vthree{m_1+m_2+m_3}
{m_1+m_2\runisymb+m_3\runisymb^2}{m_1\runisymb+m_2+m_3\runisymb^2}.
}
%\par
The projections of these vectors onto two-dimensional representation 
\eqref{S3umatsEn} are:
\Mathh{\barket{\psi} = \Vtwo{n_1+n_2\runisymb+n_3\runisymb^2}
{n_1\runisymb+n_2+n_3\runisymb^2},~~~~ 
\barket{\phi} = \Vtwo{m_1+m_2\runisymb+m_3\runisymb^2}
{m_1\runisymb+m_2+m_3\runisymb^2}.}
We discarded here the coefficient \Math{1/\sqrt{3}},
since the Born probability is a projective invariant.
Note that the vectors \Math{\psi} and \Math{\phi} vanish if and only if
\begin{equation}
n_1=n_2=n_3 \text{~~and~~} m_1=m_2=m_3,
	\label{zerocondEn} 
\end{equation}
since the primitive root of unity \Math{\runisymb} satisfies
the equation \Math{1+\runisymb+\runisymb^2 = 0}. 
Conditions \eqref{zerocondEn} determine the eigenvector of the one-dimensional
trivial subrepresentation which is orthogonal to the considered two-dimensional one.
\par
The constituents of Born's probability \eqref{BornEn} for the two-dimensional subsystem are
\begin{equation}
	\inner{\psi}{\psi}=3\vect{n_1^2+n_2^2+n_3^2}-\vect{n_1+n_2+n_3}^2,
\label{Born2den1En}	
\end{equation}
\begin{equation}
	\inner{\phi}{\phi}=3\vect{m_1^2+m_2^2+m_3^2}-\vect{m_1+m_2+m_3}^2,
\label{Born2den2En}	
\end{equation}
\begin{equation}
\cabs{\inner{\phi}{\psi}}^2=\vect{3\vect{m_1n_1+m_2n_2+m_3n_3}
-\vect{m_1+m_2+m_3}\vect{n_1+n_2+n_3}}^2.
\label{Born2numEn}	
\end{equation}
Note that: 
\begin{enumerate}
	\item Expressions \eqref{Born2den1En}--\eqref{Born2numEn} consist of the 
	\emph{invariants of permutation representation}. 
	This emphasizes the fundamental role of permutations in quantum description.
	\item 
Expressions \eqref{Born2den1En} and \eqref{Born2den2En} are always positive integer numbers.
(Unless conditions \eqref{zerocondEn}, at which these expressions vanish, hold.)
	\item
The conditions for \emph{destructive quantum interference} --- 
i.e., for vanishing Born's probability ---	are determined by the equation	
\Mathh{3\vect{m_1n_1+m_2n_2+m_3n_3}-\vect{m_1+m_2+m_3}\vect{n_1+n_2+n_3}=0.}
This equation has infinitely many solutions in natural numbers.
An example of such a solution is: 
\Math{%\set
{\barket{n} = \Vthree{1}{1}{2},~~\barket{m} = \Vthree{1}{3}{2}}}.
\end{enumerate}
Thus, we obtained, by a simple transition to the invariant subspace,
essential features of quantum behavior from ``permutation dynamics''
and ``natural'' interpretation \eqref{natamplEn} of quantum amplitude.
\section{Conclusion}
The analysis of quantum behavior with the help of finite models leads to conclusion
that quantum mechanics is --- rather then a physical theory --- 
an \emph{a priory} mathematical scheme at the heart of which the indistinguishability 
of objects lies,
i.e., this is some kind of ``\emph{calculus of indistinguishables}'' (in analogy to
the term ``calculus of infinitesimals'' of continuous mathematics).
Quantum behavior is based on the fundamental impossibility to trace
the identity of homogeneous objects in the process of their evolution.
\par
Contemplating similar subject H. Weyl wrote (\cite{WeylEn}, p. 242):
``\emph{For now we are told only how many elements, namely \Math{n_i\vect{t}}, 
are found in the state \Math{C_i\vect{t}} at any time \Math{t}, 
but no clues are available whereby to follow up the identity of 
the \Math{n} individuals through time; we do not 
know, nor is it proper to ask, whether an element that is now in the 
state, say \Math{C_5} , was a moment before in the state\Math{C_2} or \Math{C_6}.}''
\par
Therefore, the only (``statistical'') statements about numbers of certain 
invariant combinations of elements may have objective significance.
These statements must be expressed in terms of group invariants and natural numbers 
(not necessarily mutually independent) characterizing the symmetry groups, such as sizes 
of orbits, sizes of conjugacy classes, class coefficients, dimensions of irreducible
representations, etc.
\paragraph{Acknowledgment.}
The work was supported by the grants 01-01-00200 from the Russian Foundation for Basic 
Research and 3810.2010.2 from the Ministry of Education and Science of 
the Russian Federation.
\renewcommand{\refname}{References}

%---Russian part$$$$$$$$$$$$$$$$$$$$$$$$$$$$$$$$$$$$$$$$$$$$$$$$$$$$$$$$
\newpage
\def\abstractname{Аннотация}
\renewcommand{\refname}{Литература}
\setcounter{page}{1}
\setcounter{section}{0}
\setcounter{footnote}{0}
\setcounter{equation}{0}
\begin{center}
{\Large\bf
Конечные квантовые модели: конструктивный подход к описанию квантового поведения
%Finite Quantum Models: Constructive Approach to Description of Quantum Behavior
} 
\vskip 10pt 
\copyright~~
{\large\bf 
2010 г. 
{В. В. Корняк}}\\
\emph{Лаборатория информационных технологий}\\
Объединенный институт ядерных исследований\\
\emph{141980 Дубна Московской обл.}\\
\emph{E-mail: kornyak@jinr.ru}
\end{center}
%\vskip 10pt
\begin{abstract}
Универсальность квантовой механики --- её применимость к физическим системам 
совершенно различной природы и масштабов --- указывает на то, что квантовое 
поведение может быть проявлением общематематических свойств систем, 
содержащих неразличимые, т. е. лежащие на одной и той же орбите некоторой 
группы симметрий, элементы. 
В этой статье мы показываем, что квантовое поведение возникает естественным
образом в системах с конечным числом элементов связанных нетривиальными 
группами симметрий. 
``Конечный'' подход позволяет увидеть особенности квантового описания более отчетливо 
и без необходимости в концепциях типа ``коллапс волновой функции'', 
``параллельные вселенные Эверетта'' и т.~п.
В частности, в предположении конечности любая квантовая динамика 
сводится к простой перестановочной динамике. 
Преимуществом конечных квантовых моделей является возможность их конструктивного изучения 
методами компьютерной алгебры и вычислительной теории групп.
\end{abstract}

\section{Введение}
Вопрос о том, ``является ли реальный мир дискретным или непрерывным'' или даже
``конечным или бесконечным'' относится исключительно к метафизике, так как никакие эмпирические
наблюдения или логические аргументы не в состоянии обосновать тот или иной выбор ---
это вопрос веры или вкуса. Конечно, дискретные и непрерывные \emph{математические}
теории существенно различаются и эффективность их применения в физике зависит от 
конкретных исторических обстоятельств. В частности, со времён Ньютона 
и до появления современных компьютеров анализ и дифференциальная геометрия были фактически 
единственным средством математического изучения физических систем% 
\footnote{Пуанкаре подчёркивал конвенциональность выбора между дискретным 
(конечным) и непрерывным (бесконечным) описаниями природы. Поучительно проследить 
эволюцию его личных предпочтений.
%\par
В книге ``Ценность науки''
(\cite{PoincareVal}, стр. 288--289 русского перевода) он, 
отрицая фундаментальную обоснованность понятия непрерывности, высоко оценивает 
его эвристическую силу:
``\emph{Единственный естественный предмет 
математической мысли есть целое число. Непрерывность \textellipsis, без сомнения,
изобретена нами, но изобрести ее нас вынудил внешний мир.}
%\par 
\emph{Без него не было бы анализа бесконечно малых. 
Все математическое знание свелось бы к \textbf{арифметике} 
или к \label{poinc}\textbf{теории подстановок}.}
%\par 
\emph{Но мы, напротив, посвятили изучению 
непрерывности почти все наше время, почти все наши силы.}
%\par
\textellipsis\emph{Вам, без сомнения, скажут, что вне целого числа 
нет строгости, а следовательно, нет математической 
истины, что оно скрывается всюду и что нужно 
стараться разоблачить его покровы, хотя бы для этого 
пришлось обречь себя на нескончаемые повторения.}
%\par 
\emph{Но мы не будем столь строги; мы будем 
признательны непрерывности, которая, если даже \textbf{всё} 
исходит из целого числа, одна только была способна 
извлечь из него \textbf{так много}.}'' 
\par
Спустя несколько лет --- вскоре после первых наблюдений квантового 
поведения физических систем --- Пуанкаре пишет
(\cite{PoincareNew}, стр. 643 русского перевода):
``\emph{Теперь уже нельзя говорить, что 
\guillemotleft{}природа не делает скачков\guillemotright{} 
(Natura non facit saltus), --- 
на самом деле она поступает именно наоборот. 
И не только материя, возможно, сводится к атомам, 
а даже и мировая история и, я скажу, даже само 
время, поскольку два мгновения, заключенные в 
интервале между двумя скачками, не могут быть 
различимы, ибо они принадлежат одному и тому же 
состоянию мира.}''
\par
Далее Пуанкаре резюмирует сложившийся у него взгляд на проблему:
``\emph{Однако не следует слишком спешить, ибо сейчас 
очевидно лишь то, что мы весьма далеки от 
завершения борьбы между двумя стилями мышления --- одного, 
характерного для атомистов, верящих в 
существование простейших первоэлементов, очень большого, 
но конечного числа комбинаций которых 
достаточно для объяснения всего разнообразия аспектов 
Вселенной, и другого, присущего приверженцам идей 
непрерывности и бесконечности.}''}.
С развитием цифровых технологий ``дискретный'' стиль мышления становится всё более популярным,
а реальные возможности дискретной математики в приложениях существенно выросли. 
Важным преимуществом дискретного описания является его 
концептуальная ``экономность'' в оккамовском смысле --- отсутствие ``лишних сущностей'' 
основанных на идеях актуальной бесконечности типа ``дедекиндовых сечений'',
``последовательностей Коши'' и т. п. 
Более того, дискретная математика содержательно 
богаче непрерывной --- непрерывность ``сглаживает'' тонкие детали строения структур. 
Для иллюстрации этого тезиса можно сравнить списки
простых групп Ли и простых конечных групп. 
Имеется также множество аргументов в пользу того, что на малых (планковских)
расстояниях дискретное описание 
физических процессов является более адекватным и что непрерывность целесообразно 
рассматривать лишь как логическую структуру для приближённого 
(``термодинамического'') описания больших совокупностей дискретных структур.
\par
Характерной особенностью квантовой механики является её универсальность. 
Она пригодна для описания систем совершенно различной физической природы и 
размеров
 --- от элементарных частиц до больших молекул. Например, в \cite{fullerinterfer}
описаны эксперименты, в которых наблюдалась квантовомеханическая интерференция
между молекулами фуллерена \Math{C_{60}}.
Такой универсальностью обычно обладают теории, в основе которых заложены некие априорные
математические принципы. Примером подобной универсальной математической 
схемы является \emph{статистическая механика}. В её основе лежат не зависящие от конкретной 
физической системы принципы, главными из которых являются классификация микросостояний 
по уровням энергии и постулат равномерного распределения энергии по степеням свободы. 
В случае квантовой механики ведущим математическим принципом является симметрия: 
квантовомеханическое поведение демонстрируют системы, содержащие неразличимые частицы 
--- любое нарушение тождественности частиц разрушает квантовые интерференции.
\par
Мы рассматриваем здесь основные структуры квантового описания предполагая
конечность всех входящих в формулировки множеств. При таком подходе, 
перефразируя Пуанкаре, 
``всё можно свести к арифметике и теории подстановок''.
%_____________________________________________________________________
\section{Основные конструкции и обозначения}
\subsection{Классическая и квантовая эволюция}
Мы рассматриваем эволюцию \emph{динамической системы} с конечным множеством
\emph{состояний} \Math{\wS=\set{\ws_1,\ldots,\ws_{\wSN}}}
в дискретном \emph{времени} \Math{t\in\Time=\set{\ldots,-1,0,1,\ldots}}.
Рассматривая \emph{конечные} эволюции мы можем предположить для упрощения обозначений, что 
\Math{\Time=\ordset{\tin,1,\ldots,\tfin}}, где \Math{\tfin\in\Nat}.
\par
\emph{Классическая эволюция} (\emph{история}, \emph{траектория}) динамической системы 
это последовательность состояний в зависимости от времени 
\Math{\ldots,s_{t-1},s_{t},s_{t+1},\ldots\in\wS^\Time}.
\par
Мы предполагаем, что на множестве состояний действует конечная 
\emph{группа симметрий}
\Math{\wG=\set{\wg_1=\id,\ldots,\wg_{\wGN}}\leq\Perm{\wS}}.
\par
\emph{Квантовой эволюцией} мы будем называть последовательность перестановок 
\Math{\ldots{}p_{t-1},p_{t},p_{t+1}\ldots\in\wG^\Time},
определяющую произведение \Math{\cdots{}p_{t-1}\,p_{t}\,p_{t+1}\cdots\in\wG}.
Смысл такого определения будет ясен из дальнейшего.
%\subsection{Динамические системы с пространством}
\subsection{Динамические системы с пространственной структурой}
В физике полное множество состояний \Math{\wS} имеет, как правило, специальную 
структуру множества функций \Math{\wS=\lSX} на некотором \emph{пространстве} 
\Math{\X} со значениями
в некотором множестве \emph{локальных состояний} \Math{\lS}.
В динамических системах с пространством естественным образом возникают нетривиальные
калибровочные
структуры, использующиеся в физических теориях для описания 
силовых полей. 
\par
Мы предполагаем, что \emph{пространство} представляет собой конечное множество
\Math{\X=\set{\x_1,\ldots, \x_\XN}} симметрии которого образуют 
\emph{группу пространственных симметрий} \Math{\sG=\set{\sg_1=\id,\ldots,\sg_\sGN}\leq\Perm{\X}}.
Случай, когда \Math{\sG} является \emph{собственной} подгруппой группы \Math{\Perm{\X}}
всех возможных перестановок точек из \Math{\X},
подразумевает наличие дополнительной структуры у множества \Math{\X}. Например
 --- и этого достаточно для наших целей --- \Math{\X} может быть
\emph{абстрактным} графом.
\par
\emph{Локальные состояния} образуют конечное множество
\Math{\lS=\set{\ls_1,\ldots,\ls_\lSN}} с группой \emph{внутренних симметрий}
\Math{\iG=\set{\ig_1=\id,\ldots, \ig_\iGN}\leq\Perm{\lS}}.
\par
Группа \Math{\wG} симметрий полного множества состояний \Math{\wS} может
быть построена из пространственных \Math{\sG} и внутренних \Math{\iG}
симметрий различными способами. Естественным обобщением конструкций, используемых
в физических теориях является представление \Math{\wG} в виде следующего расщепляемого
расширения
\begin{equation}
\id\rightarrow\iGX\rightarrow\wG\rightarrow\sG\rightarrow\id,
\label{extention}
\end{equation}
где \Math{\iGX} --- группа \Math{\iG}-значных функций на пространстве \Math{\X}.
Явные формулы, выражающие групповые операции в 
\Math{\wG} из \eqref{extention} в терминах операций в группах \Math{\sG} и \Math{\iG}, приведены в
\cite{KornyakJMS10, KornyakNS10} --- в данной статье они нам не понадобятся.
%______________________________________________________________
\subsection{Некоторые обозначения}
Имея в виду дальнейшие цели, нам удобно включать нуль во множество натуральных чисел,
т.е. мы будем использовать определение: \Math{\Nat=\set{0,1,2,\ldots}}.
\par
Если необходимо явно указать является ли элемент \Math{\psi} гильбертова пространства 
\Math{\Hspace} вектором или ковектором%
\footnote{Гильбертово пространство ввиду наличия скалярного произведения 
канонически изоморфно своему дуальному пространству.}
будут использоваться обозначения \Math{\barket{\psi}} или \Math{\brabar{\psi}}, соответственно.
\par 
Для \emph{стандартного скалярного произведения} в \Math{\adimH}-мерном гильбертовом 
пространстве мы используем круглые скобки: 
\begin{equation}
	\innerstandard{\phi}{\psi}=\sum\limits_{i=1}^{\adimH}\cconj{\phi^i}\psi^i.
\label{innerstd}
\end{equation}
\par
Для \emph{инвариантного скалярного произведения} используются угловые скобки:
\begin{equation}
\inner{\phi}{\psi}=\frac{\textstyle{1}}{\textstyle{\cabs{G}}}\sum\limits_{g\in{}G}
\!\innerstandard{U\!\vect{g}\phi}{U\vect{g}\psi},
\label{innerinv}
\end{equation}
где \Math{U} --- представление группы \Math{G} в гильбертовом пространстве \Math{\Hspace}.

\section{Квантовая эволюция динамической системы}
Наиболее популярным и интуитивно ясным методом квантования является фейнмановское квантование 
с помощью интегралов по траекториям \cite{Feynman}. Этот метод особенно удобен для динамических
систем с пространственной структурой. Согласно подходу Фейнмана амплитуда квантового перехода
системы из начального состояния в конечное вычисляется с помощью суммирования амплитуд вдоль 
всех возможных классических траекторий, соединяющих эти состояния. 
Амплитуда вдоль отдельной траектории вычисляется как произведение амплитуд переходов между
ближайшими последовательными состояниями на траектории. 
Стандартно амплитуда имеет вид экспоненты от действия вдоль траектории
\begin{equation}
A_{\mathrm{U}(1)}~=~A_0\exp\vect{iS}=A_0\exp\vect{i\int\limits_0^T{}Ldt}.
\label{amplclass}
\end{equation}
Функция \Math{L}, зависящая от {первых} производных состояний по времени, называется
\emph{лагранжианом}. В дискретном времени экспонента от интеграла переходит
в произведение \Math{\exp\vect{i\int{}Ldt}\rightarrow
	\e^{{iL_{0,1}}}\ldots\e^{{iL_{t-1,t}}}
	\ldots\e^{{iL_{T-1,T}}}} и выражение для амплитуды принимает вид
\begin{equation}
	A_{\mathrm{U}(1)}~=~A_0
	\e^{{iL_{0,1}}}\ldots\e^{{iL_{t-1,t}}}
	\ldots\e^{{iL_{T-1,T}}}.\label{ampldiscr}
\end{equation}
Элементы \Math{\Partransport_{t-1,t}=\e^{{iL_{t-1,t}}}} этого произведения 
естественно интерпретировать как \emph{связности} (\emph{параллельные переносы}) со значениями
в \emph{одномерном} унитарном представлении окружности, т. е. коммутативной группы Ли
\Math{\iG=S^1\equiv\R/\Z}.
\par 
Естественное обобщение возникает из предположений, что
группа \Math{\iG} не обязательно окружность и что её унитарное представление 
\Math{\Rep{\iG}} не обязательно одномерно.	
В этом случае амплитуда представляет собой многокомпонентный вектор, что удобно для описания
частиц с внутренними степенями свободы. Значение такой многокомпонентной амплитуды на
 траектории принимает вид%
\footnote{В некоммутативном случае необходимо соблюдать порядок операторов, согласованный
в данном случае с традицией писать матрицы слева от векторов.}
\begin{equation}
	A_{\Rep{\iG}}=\Rep{\alpha_{T,T-1}}\ldots\Rep{\alpha_{t,t-1}}
	\ldots\Rep{\alpha_{1,0}}A_0,\hspace*{10pt}\alpha_{t,t-1}\in\iG.\label{amplgen}
\end{equation}
Мы будем предполагать, что \Math{\iG} --- конечная группа. Линейные представления конечных 
групп автоматически унитарны. Ясно, что стандартное квантование \eqref{amplclass} можно 
аппроксимировать с помощью одномерных представлений конечных циклических групп.
\par
Хорошо известно, что фейнмановский подход эквивалентен традиционной матричной формулировке
квантовой механики. В этой формулировке эволюция системы из начального состояния в конечное 
описывается матрицей эволюции \Math{U}:
\Math{\barket{\psi_{\tin}}\rightarrow\barket{\psi_{\tfin}}=U\barket{\psi_{\tin}}}.
Матрица эволюции может быть представлена в виде произведение матриц, соответствующих
элементарным шагам во времени: 
\Mathh{U=U_{\tfin\leftarrow\tfin-1}\cdots{}U_{t\leftarrow{}t-1}\cdots{}U_{1\leftarrow0}.}
Фактически, фейнмановские правила квантования ``перемножать последовательные амплитуды''
и ``суммировать альтернативные истории'' --- это изложение другими словами правил умножения
матриц. Это ясно из иллюстрации, на которой два шага эволюции квантовой системы с двумя 
состояниями (однокубитный регистр) представлены параллельно в фейнмановской и матричной формах:
\begin{center}
\begin{tabular}[t]{ccc}
\includegraphics[height=0.24\textwidth]{Feynman-MatrixBA}
&\raisebox{0.1\textwidth}{$\sim$}%{$\Huge\Longleftrightarrow$}
&
\includegraphics[height=0.24\textwidth]{Feynman-MatrixU}
\\
$\Updownarrow$&&$\Updownarrow$
\\
$
BA=
\begin{bmatrix}
{\color{blue}b_{11}a_{11}+b_{12}a_{21}}
&
{\color{red}b_{11}a_{12}+b_{12}a_{22}}
\\[5pt]
{\color{blue}b_{21}a_{11}+b_{22}a_{21}}
&
{\color{blue}b_{21}a_{12}+b_{22}a_{22}}
\end{bmatrix}
$~~~~~~~~~
& $\sim$~~ &
$
U=
\begin{bmatrix}
{\color{blue}u_{11}}&{\color{red}u_{12}}
\\[5pt]
{\color{blue}u_{21}}&{\color{blue}u_{22}}
\end{bmatrix}
$~~~~~~~~
\end{tabular}
\end{center}
В соответствии с фейнмановскими правилами переход, скажем, между состояниями 
\Math{\phi_2} и \Math{\psi_1} определяется суммой по двум путям
 \Math{b_{11}a_{12}+b_{12}a_{22}}.
Но это же выражение является элементом \Math{u_{12}} произведения матриц \Math{U=BA}.
Общий случай произвольного числа состояний и произвольного числа шагов по времени
легко выводится по индукции из этого элементарного примера.
\par
Это рассуждение применимо
и в случае некоммутативной связности, когда амплитуды вдоль путей описываются формулой
\eqref{amplgen}. Нам достаточно лишь трактовать эволюционные матрицы
\Math{A, B} и \Math{U} как блочные матрицы с некоммутативными элементами,
являющимися матрицами из представления \Math{\Rep{\iG}}.
Для единообразия рассмотрения мы можем проигнорировать блочную структуру матриц 
интерпретируя их как обычные матрицы большей размерности из представлений группы 
полных симметрий \Math{\wG}, 
сконструированной в соответствии с \eqref{extention}.
\par
В квантовой механике, эволюционные матрицы \Math{U} представляют собой унитарные операторы,
действующие в гильбертовых пространствах \emph{векторов состояний} (называемых также
``\emph{волновыми функциями}'', ``\emph{амплитудами}'' и т.д.).
\emph{Квантовомеханические
частицы} ассоциируются с унитарными представлениями определённых групп. Эти представления,
в соответствии с их размерностями, называются ``\emph{синглетами}'', 
``\emph{дублетами}'' и т.д. Многомерные представления описывают \emph{спин}.
Квантовомеханический эксперимент сводится к сравнению вектора состояния системы
\Math{\psi} с некоторым эталонным вектором состояния \Math{\phi},
обеспечиваемым ``\emph{измерительным аппаратом}''. 
В соответствии с правилом Борна, вероятность наблюдать
совпадение состояний равна \Math{{\cabs{\inner{\phi}{\psi}}^2}} 
(в предположении нормировки \Math{\inner{\phi}{\phi}=\inner{\psi}{\psi}=1}). 
Чтобы обеспечить конструктивность всех этих понятий квантовой механики мы 
предполагаем далее, что операторы эволюции являются элементами представлений конечных
групп. 
\section{Квантовое описание конечных систем}
\subsection{Перестановки и линейные представления}
\subsubsection{Действия группы.}
Легко описать \cite{Hall} все транзитивные действия конечной группы
\Math{\wG=\set{\wg_1,\ldots,\wg_{\wGN}}} 
на конечных множествах \Math{\Omega=\set{\omega_1,\ldots,\omega_n}}.
Любое такое множество находится во взаимно однозначном соответствии с \emph{правыми}
(или \emph{левыми}) смежными классами по некоторой подгруппе \Math{H\leq\wG}, т. е.
\Math{\Omega\cong{}H\backslash\wG} (или \Math{\Omega\cong{}\wG/H}).
Множество \Math{\Omega} называется \emph{однородным пространством} группы \Math{\wG} 
(\Math{\wG}-\emph{пространством}).
Действие \Math{\wG} на \Math{\Omega} является \emph{точным}, если подгруппа
\Math{H} не содержит нормальных подгрупп группы \Math{\wG}. 
Мы можем написать действие в виде перестановок
\Mathh{
\pi(g)=\dbinom{\omega_i}{\omega_ig}\sim\dbinom{Ha}{Hag},
\hspace*{20pt}g,a\in{}\wG,~~~i=1,\ldots,n.
}
\par
Максимальным транзитивным множеством \Math{\Omega} является множество всех элементов 
самой группы \Math{\wG}, т. е. смежных классов по тривиальной подгруппе \Math{H=\set{\id}}.
Соответствующее действие называется \emph{регулярным} и может быть 
представлено перестановками
\begin{equation}
	\Pi(g)=\dbinom{\wg_i}{\wg_ig},~~~~i=1,\ldots,\wGN.
	\label{regperm}
\end{equation}
\par
Для того чтобы ввести ``количественное'' (``статистическое'') описание предпишем 
элементам множества \Math{\Omega} числовые ``веса'' из какой-нибудь подходящей
 \emph{числовой системы} \Math{\aNumbers}, содержащей, по крайней мере, нуль и единицу.
В этом случае перестановки можно переписать в матричной форме 
\begin{equation}
\pi(g)\rightarrow\rho(g)=
\Mone{\rho(g)_{ij}},\text{~~ где~~} \rho(g)_{ij}=\delta_{\omega_ig,\omega_j};~~ i,j=1,\ldots,n;
\label{permrep}
\end{equation}
\Mathh{\delta_{\alpha,\beta}\equiv
\begin{cases}
1, & \text{если~~} \alpha=\beta,\\
0, & \text{если~~} \alpha\neq\beta,
\end{cases}
\text{~~для~~} \alpha,\beta\in\Omega.
}
Функция \Math{\rho}, определённая формулой \eqref{permrep}, называется
\emph{перестановочным представлением}.
\par
\emph{Цикловым типом} перестановки называется массив кратностей длин 
циклов в разложении перестановки на непересекающиеся циклы. 
Этот массив обычно записывается в виде
\Math{1^{k_1}2^{k_2}\cdots{}n^{k_n},}
где \Math{k_i} --- число циклов длины \Math{i} в перестановке; 
\Math{0\leq{}k_i\leq{}n;~~k_1+2k_2+\cdots+nk_n=n.}
Исходя из циклового типа перестановки \Math{\pi(g)} нетрудно выписать 
{характеристический многочлен} 
перестановочной матрицы \eqref{permrep}:
\begin{equation}
\chi_{\rho(g)}\vect{\lambda}=\det\vect{\rho(g)-\lambda\idmat}
=\vect{\lambda-1}^{k_1}\vect{\lambda^2-1}^{k_2}\cdots\vect{\lambda^n-1}^{k_n}.
\label{charpol}	
\end{equation}
Матричная форма \emph{регулярного} действия \eqref{regperm} 
\begin{equation}
	\Pi(g)\rightarrow{}\regrep(g)=
\Mone{\regrep(g)_{ij}},~~ \regrep(g)_{ij}=\delta_{e_ig,e_j},~~ i,j=1,\ldots,\wGN
	\label{regrep}
\end{equation}
называется \emph{регулярным представлением} --- это частный случай 
перестановочного представления \eqref{permrep}.
\par
Для свободы алгебраических манипуляций обычно 
предполагается, что \Math{\aNumbers} --- алгебраически замкнутое поле, например, поле
комплексных чисел \Math{\C}. 
%Более экономной числовой системой, используемой в теории
%конечных групп, является расширение поля рациональных чисел с помощью 
%примитивного корня из единицы: \Math{\aNumbers=\Q\vect{\runi{}}}.
Если \Math{\aNumbers} --- поле, то множество \Math{\Omega} можно
рассматривать как базис линейного векторного пространства 
	\Math{\Hspace=\mathrm{Span}\vect{\omega_1,\cdots,\omega_n}}.
\subsubsection{Системы чисел} 
Из \eqref{charpol} видно, что все \emph{собственные значения} перестановочных матриц 
являются корнями из единицы. Этот факт --- в сочетании с тем, что все неприводимые
представления конечных групп являются подпредставлениями регулярного представления
\eqref{regrep} --- означает, что все числа, достаточные для
наших целей можно построить из множества натуральных чисел \Math{\N=\set{0,1,\ldots}} 
и примитивного корня из единицы \Math{\runi{}} некоторой степени \Math{n}. 
Термин \emph{примитивный} означает, что \Math{\runi{}^n=1} и период \Math{\runi{}} 
равен в точности \Math{n}.
В роли \Math{n} всегда можно использовать \emph{экспоненту} группы 
--- наименьшее общее кратное
порядков элементов группы. Но часто бывает достаточно некоторого собственного делителя
экспоненты.
Любой корень из единицы имеет вид \Math{\runi{}^k,~k\in\set{0,1,\ldots,n-1}}. 
Для наглядности можно считать символически, что \Math{\runi{}=\e^{2\pi{}i/n}}, 
но такое представление нам нигде не понадобится. 
Достаточно только следующих \emph{алгебраических} определений
\begin{enumerate}
	\item \Math{\runi{}^k\times\runi{}^m=\runi{}^{k+m\mod{}n}} --- \emph{правило умножения},
	\item \Math{\cconj{\runi{}^k}=\runi{}^{n-k}} --- \emph{комплексное сопряжение}.
\end{enumerate}
Таким образом, в качестве \emph{системы чисел}, мы будем использовать полиномы от
\Math{\runi{}} с \emph{натуральными} коэффициентами:
 \Math{\aNumbers=\N\ordset{\runi{}}}.
\par
Если \Math{n=1}, то \Math{\aNumbers} представляет собой
\emph{полукольцо натуральных чисел} \Math{\N}. 
Эта система чисел соответствует случаю \emph{тривиальной} группы \Math{\wG}.
\par
При \Math{n\geq2} можно ввести \emph{отрицательные числа}
с помощью определения
\Mathh{\vect{-1}=
\begin{cases}
\runi{}^{n/2}, & \text{если~~}n\text{~~чётное},\\
\runi{}+\cdots+\runi{}^{n-1}, & \text{если~~}n\text{~~нечётное.}
\end{cases}
}
В частности, при \Math{n=2} возникает \emph{кольцо целых чисел}: \Math{\aNumbers=\Z}.
Эта система чисел соответствует группе \Math{\wG=\CyclG{2}} 
\Math{\vect{\text{или~}\wG=\CyclG{2}\times\cdots\times\CyclG{2}}}. 
\par
При \Math{n\geq3} множество 
\Math{\aNumbers=\N\ordset{\runi{}}} представляет собой \emph{коммутативное кольцо}, 
которое можно погрузить в поле комплексных чисел \Math{\C}.
\par
Кольца \Math{\N\ordset{\runi{}}} достаточно для проведения всех вычислений с 
конечными квантовыми моделями. 
Для упрощения изложения, чтобы можно было свободно говорить о линейных пространствах, 
мы будем использовать также поле частных%
\footnote{С вычислительной точки зрения работа с кольцами, 
хотя и слегка усложняет алгоритмы, более эффективна, чем вычисления 
над соответствующими полями частных.
Современные системы компьютерной алгебры, как правило, 
используют вычисления над кольцами в алгоритмах линейной алгебры.}
этого кольца. 
Иногда, как промежуточные технические символы, могут возникать 
квадратные корни из натуральных чисел (которые, в принципе, 
также можно выразить через
корни из единицы, но в этом нет реальной необходимости).
В окончательных выражениях, имеющих статус ``наблюдаемых'', 
ни корни из единиц, ни другие иррациональности не появляются.
%___________________________________________________________________
\subsubsection{Унитарные представления}
Унитарные операторы играют ключевую роль в квантовой механике.
Любое линейное представление конечной группы эквивалентно унитарному, поскольку
всегда можно сконструировать инвариантное скалярное произведение 
из произвольного ``усреднением по группе''. 
Например, инвариантное произведение \eqref{innerinv}
построено усреднением стандартного \eqref{innerstd}.
\par
Изложим кратко основные сведения о неприводимых представлениях конечных групп \cite{Hall}, 
иллюстрируя их с помощью наименьшей некоммутативной группы --- группы перестановок 
трёх элементов \Math{\SymG{3}}. Эта группа изоморфна также группе симметрий 
треугольника, т.е. диэдральной группе \Math{\DihG{6}\cong\SymG{3}}.
Группа состоит из шести элементов, имеющих следующее представление в виде перестановок
\begin{equation}
\wg_1=\vect{},~\wg_2=\vect{2,3},~\wg_3=\vect{1,3},~\wg_4=\vect{1,2},
	~\wg_5=\vect{1,2,3},~\wg_6=\vect{1,3,2}.
\label{S3elems}	
\end{equation}
\par
Аналогом замены системы координат в физике в теории групп является сопряжение: 
\Math{a^{-1}ga\rightarrow{}g',} \Math{g, g'\in\wG,} \Math{a\in\Aut{\wG}}.
Сопряжения элементами самой группы, т.е. когда \Math{a\in\wG},
называются \emph{внутренними автоморфизмами}. Классы эквивалентности элементов группы
относительно внутренних автоморфизмов называются \emph{классами сопряжённых элементов}
(или, короче, \emph{классами сопряжённостей}). 
Разложение группы на классы сопряжённостей, символически записываемое в виде
\Mathh{\wG=\class{1}+\class{2}+\cdots+\class{\classN},} 
играет центральную роль в изучении её представлений.
\par
\emph{Пример.} Группа \Math{\SymG{3}} распадается на три класса сопряжённостей:
\begin{equation}
	\class{1}=\set{\vect{}},~~\class{2}=\set{\vect{2,3},~\vect{1,3},~\vect{1,2}},
	~~\class{3}=\set{\vect{1,2,3},~\vect{1,3,2}}.
	\label{S3classes}
\end{equation}
\par
Умножение в группе позволяет ввести операцию умножения для классов:
\emph{произведением классов} \Math{\class{i}} и \Math{\class{j}}
называется разложенное на классы \emph{мультимножество} всех возможных
произведений \Math{ab,~a\in\class{i},~b\in\class{j}}. 
Очевидно, что так определённое произведение коммутативно,
поскольку \Math{ab} и \Math{ba} входят в один и тот же класс:
\Math{ab\sim{}a^{-1}\vect{ab}a=ba}. 
Таким образом, таблица умножения классов имеет вид
\begin{equation}
	\class{i}\class{j}= \class{j}\class{i} = \sum\limits_{k=1}^{\classN}c_{ijk}\class{k}.
\label{classtab}	
\end{equation}
\emph{Натуральные} целые числа \Math{c_{ijk}}
--- кратности классов в соответствующих мультимножествах ---
называются \emph{коэффициентами (алгебры) классов}.
\par
\emph{Пример.} Группа \Math{\SymG{3}} имеет следующую таблицу умножения классов
\Mathh{\class{1}\class{j}=\class{j},~~\class{2}^2=3\class{1}+3\class{3},
~~\class{2}\class{3}=2\class{2},~~\class{3}^2=2\class{1}+\class{3}.}
\par
\textit{Краткий список основных свойств линейных представлений конечных групп:}
\begin{enumerate}
	\item 
Каждое линейное представление унитарно (эквивалентно унитарному).
	\item 
Любое неприводимое представление содержится в регулярном. 
Более конкретно,	существует (унитарная) матрица \Math{\transmatr} одновременно 
приводящая все матрицы \eqref{regrep} к виду
\begin{equation}
	\transmatr^{-1}\regrep(g)\transmatr=
	\bmat
			\repirr_1(g) &&&
	\\[5pt]
	&
	\hspace*{-27pt}d_2\left\{
	\begin{matrix}
	\repirr_2(g)&&\\
	&\hspace*{-10pt}\ddots&\\
	&&\hspace*{-7pt}\repirr_2(g)
	\end{matrix}
	\right. 
	&&
	\\
	&&\hspace*{-10pt}\ddots&\\
	&&& 
	\hspace*{-25pt}d_{\classN}\left\{
	\begin{matrix}
	\repirr_{\classN}(g)&&\\
	&\hspace*{-10pt}\ddots&\\
	&&\hspace*{-7pt}\repirr_{\classN}(g)
	\end{matrix}
	\right. 
	\emat
\label{regrepdecomp}	
\end{equation}
и любое неприводимое представление является одним из элементов множества
\Math{\set{\repirr_1,\ldots,\repirr_{\classN}}}.
Число неэквивалентных неприводимых представлений
\Math{\classN} равно числу классов сопряжённостей \Math{\class{j}} группы \Math{\wG}.
Число \Math{d_j} это одновременно размерность неприводимого представления \Math{\repirr_j}
и кратность его вхождения в регулярное представление. 
Из \eqref{regrepdecomp} видно, что для размерностей \Math{d_j}
выполняется соотношение: 
\Math{d^2_1+d^2_2+\cdots+d^2_{\classN}=\cabs{\wG}=\wGN.}
Можно показать также, что размерности делят порядок группы:
\Math{d_j\mid\wGN.}
	\item 
Каждое неприводимое представление \Math{\repirr_j} однозначно (с точностью до изоморфизма)
определяется своим \emph{характером} \Math{\chi_j}, т. е. функцией на группе, определяемой
как след матрицы представления: \Math{\chi_j\vect{g}=\mathrm{Tr}\repirr_j\vect{g}}.
Фактически характер является функцией на классах сопряжённостей поскольку
\Math{\chi_j\vect{g}=\chi_j\vect{a^{-1}ga}}. Функции на классах называются центральными
\emph{центральными}.
Любая центральная функция является линейной комбинацией характеров. Значение характера 
\Math{\chi_j} на классе \Math{\class{1}=\set{\id}} равно размерности представления \Math{d_j}.
	\item 
Компактной формой регистрации	всех неприводимых представлений 
%\Math{\wGN} 
является \emph{таблица характеров.} 
Столбцы этой таблицы пронумерованы классами сопряжённых
элементов, 
а строки содержат значения характеров неэквивалентных представлений:
\begin{center}
\begin{tabular}{c|cccc}
&\Math{\class{1}}&\Math{\class{2}}&\Math{\cdots}&\Math{\class{\classN}}\\\hline
\Math{\chi_1}&1&1&\Math{\cdots}&1\\
\Math{\chi_2}&\Math{\chi_2\vect{\class{1}}=d_2}&\Math{\chi_2\vect{\class{2}}}&
\Math{\cdots}&\Math{\chi_2\vect{\class{\classN}}}\\
\Math{\vdots}&\Math{\vdots}&\Math{\vdots}&&\Math{\vdots}\\
\Math{\chi_{\classN}}&\Math{\chi_{\classN}\vect{\class{1}}=d_{\classN}}&
\Math{\chi_{\classN}\vect{\class{2}}}&\Math{\cdots}&\Math{\chi_{\classN}\vect{\class{\classN}}}
\end{tabular}.
\end{center}
В соответствии со стандартным соглашением, первый столбец соответствует классу 
единичного элемента группы, а первая строка содержит одномерное
 \emph{тривиальное} представление.
\end{enumerate}

%_____________________________________________________________
\subsection{Погружение квантовой системы в классическую}
В самой общей формулировке квантовая механика предполагает, что каждой физической
системе соответствует гильбертово пространство \Math{\Hspace}
ненулевые векторы которого, \Math{\psi\in\Hspace}, представляют все возможные состояния
системы. Предполагается, что векторы \Math{\psi} и \Math{\psi'} описывают одно и то же
состояние, если \Math{\psi'=\lambda\psi,~ \lambda\in\C.}
Эволюция системы из состояния \Math{\psi_0} в состояние \Math{\psi_T}
описывается \emph{унитарным} оператором \Math{U}: 
\Math{\barket{\psi_T}=U\barket{\psi_0}}. 
 Унитарность означает, что \Math{U} является 
элементом группы автоморфизмов пространства \Math{\Hspace}:
\Math{U\in\Aut{\Hspace}}. Можно считать, что \Math{\Aut{\Hspace}}
является точным унитарным представлением соответствующей абстрактной группы \Math{\wG}.
В непрерывном времени динамику можно выразить в терминах локального 
\emph{оператора энергии} 
(\emph{гамильтониана}) \Math{H} с помощью уравнения Шрёдингера
\Mathh{i\frac{\mathrm{d}}{\mathrm{d}t}\barket{\psi}=H\barket{\psi}.}
Если \emph{эрмитов} оператор \Math{H} не зависит от времени, то он связан с оператором эволюции
простым соотношением \Math{U=\e^{-iHT}.}
\par
Конечная квантовая система строится точно по такой же схеме. 
Только теперь группа \Math{\wG} --- конечная группа порядка \Math{\wGN}.
Все возможные операторы эволюции образуют конечное множество 
\Math{\set{U_1,\ldots,U_{\wGN}}}
матриц унитарного представления \Math{\mathrm{U}} в 
\Math{\adimH}-мерном гильбертовом пространстве \Math{\Hspace_{\adimH}}. Поскольку матрицы \Math{U_j} 
невырожденные, всегда можно
ввести гамильтонианы по формуле \Math{H_j=i\ln{}U_j}, но в этом нет никакой необходимости.
Заметим, что гамильтонианы практически не используются во многих приложениях квантовой механики, 
например, в \emph{квантовых вычислениях} и в физических теориях, основанных на 
\emph{матрице рассеяния} (\Math{S}-\emph{матрице}).
\par
Конечные группы, если они ``достаточно некоммутативны'', 
зачастую порождаются небольшим 
числом элементов.
Например, все простые и все симметрические группы порождаются двумя элементами.
Алгоритм построения всей группы, исходя из \Math{n_{g}} порождающих элементов, 
весьма прост и сводится к \Math{n_{g}\vect{\wGN-n_{g}-1}} групповым умножениям.
Таким образом, квантовая динамика конечных систем достаточно удобна для
исследования методами компьютерной алгебры. 
\par
Из разложения \eqref{regrepdecomp} видно, 
что мы \emph{всегда} можем расширить \Math{\adimH}-мерное представление \Math{\mathrm{U}} до
\Math{\wSN}-мерного представления \Math{\mathrm{\widetilde{U}}} 
в гильбертовом пространстве \Math{\Hspace_{\wSN}}, эквивалентного представлению, 
соответствующему перестановкам \emph{некоторого} \Math{\wSN}-элементного
множества состояний \Math{\wS=\set{\ws_1,\ldots,\ws_{\wSN}}}. Ясно, что \Math{\wSN\geq\adimH}.
\par
Ситуация, когда \Math{\wSN} строго больше чем \Math{\adimH} наиболее интересна.
Ясно, что дополнительные ``\emph{скрытые параметры}'' --- появляющиеся в этом случае из-за
увеличения числа состояний (размерности пространства) --- никоим образом не могут повлиять
на данные, относящиеся к пространству \Math{\Hspace_{\adimH}}, поскольку как само
\Math{\Hspace_{\adimH}}, так и его дополнение в \Math{\Hspace_{\wSN}}
являются инвариантными подпространствами в расширенном пространстве \Math{\Hspace_{\wSN}}.
Таким образом, мы можем \emph{любую квантовую
задачу} в \Math{\adimH}-мерном гильбертовом пространстве свести к перестановкам 
 \Math{\wSN} элементов.
\par 
Алгоритмически операции с перестановками намного более эффективны, 
чем работа с матрицами по правилам линейной алгебры,
но, с другой стороны, степени \Math{\wSN} перестановок могут существенно превосходить 
размерности \Math{\adimH} матриц. Впрочем, с идейной точки зрения принципиальная возможность 
\emph{сведения квантовой эволюции к перестановкам} гораздо важнее алгоритмических вопросов.
\par
\paragraph
{Пример. Группа \Math{\SymG{3}}}
 имеет следующую таблицу характеров 
\begin{equation}
	\text{\begin{tabular}{c|crr}
	&\Math{\class{1}}&\Math{\class{2}}&\Math{\class{3}}\\\hline
	\Math{\chi_1}&1&1&1\\
	\Math{\chi_2}&1&-1&1\\
	\Math{\chi_3}&2&0&-1
	\end{tabular}\enspace.}\label{S3tab}
\end{equation}
Напомним, что элементный состав классов \Math{\class{j}} выписан в \eqref{S3classes}.
В качестве представления \Math{\mathrm{U}}, описывающего эволюции квантовой системы,
выберем двумерное точное представление, соответствующее характеру
\Math{\chi_3}. Матрицы (операторы эволюции) этого представления
в соответствии с порядком, в котором элементы группы %\Math{\SymG{3}} 
перечислены в \eqref{S3elems}, имеют вид
\begin{align}
		U_1=\Mtwo{1}{0}{0}{1},~U_2=\Mtwo{0}{\runi{}^2}{\runi{}}{0},
		~U_3=\Mtwo{0}{\runi{}}{\runi{}^2}{0},\nonumber\\[-6pt]
		\label{S3umats}\\[-6pt]
		~U_4=\Mtwo{0}{1}{1}{0},
		~U_5=\Mtwo{\runi{}^2}{0}{0}{\runi{}},~U_6=\Mtwo{\runi{}}{0}{0}{\runi{}^2}.\nonumber
	\end{align}
Здесь \Math{\runi{}} --- примитивный корень третьей степени из единицы.%
\footnote{Обратим внимание на характерную особенность
представления \eqref{S3umats} --- его матрицы по структуре очень похожи на матрицы 
перестановок: в каждом столбце и каждой строке имеется ровно один ненулевой элемент,
только вместо \emph{единиц}, как в матрицах перестановок, здесь ненулевые элементы 
--- \emph{корни из единицы}. 
Эта особенность является следствием того, что группа \Math{\SymG{3}}
относится к так называемым \emph{\textbf{мономиальным группам}} у которых все неприводимые 
представления можно построить как индуцированные с одномерных представлений некоторых подгрупп
(в случае \eqref{S3umats} такой подгруппой является \Math{\CyclG{3}\leq\SymG{3}}).
Заметим, что, по крайней мере, для невысоких порядков, большинство групп являются именно
мономиальными. Так, например, число всех неизоморфных групп порядка, меньшего 384, 
равно 67424.
Из них только 249 групп являются \emph{\textbf{немономиальными}}. 
Наименьшей немономиальной группой является 24-элементная группа 
\Math{\SL{2}{3}} матриц размера \Math{2\times2} 
в характеристике 3 с единичным определителем.
}
%\paragraph{Замечание. Мономиальные и немономиальные группы.} 
\paragraph{Квантовые и перестановочные матрицы для \Math{\SymG{3}}.} 
Поскольку любое перестановочное представление всегда содержит одномерное 
инвариантное подпространство
порождаемое вектором \Math{\Vthree{1}{\vdots}{1}}, есть только один способ расширить
\eqref{S3umats} до перестановок трёх элементов --- необходимо добавить тривиальное 
представление, соответствующее характеру \Math{\chi_1} из таблицы \eqref{S3tab}.
Таким образом, мы приходим к трёхмерному представлению \Math{\mathrm{\widetilde{U}}}
матрицы которого имеют вид
\begin{equation}
	\widetilde{U}_j=\Mtwo{1}{0}{0}{U_j},~~j =1,\ldots,6.
	\label{S3permq}
\end{equation}
Эти матрицы представляют собой перестановочные матрицы в базисе, в котором 
перестановочное представление разложено на инвариантные компоненты. 
Мы будем называть такой базис \emph{квантовым}.
В \emph{перестановочном} базисе эти матрицы имеют вид
\begin{align}
		P_1=\Mthree{1}{\cdot}{\cdot}{\cdot}{1}{\cdot}{\cdot}{\cdot}{1},
		~P_2=\Mthree{1}{\cdot}{\cdot}{\cdot}{\cdot}{1}{\cdot}{1}{\cdot},
		~P_3=\Mthree{\cdot}{\cdot}{1}{\cdot}{1}{\cdot}{1}{\cdot}{\cdot},\nonumber\\[-6pt]
		\label{S3pmats}\\[-6pt]
		~P_4=\Mthree{\cdot}{1}{\cdot}{1}{\cdot}{\cdot}{\cdot}{\cdot}{1},
		~P_5=\Mthree{\cdot}{1}{\cdot}{\cdot}{\cdot}{1}{1}{\cdot}{\cdot},
		~P_6=\Mthree{\cdot}{\cdot}{1}{1}{\cdot}{\cdot}{\cdot}{1}{\cdot}.\nonumber
	\end{align}
Наиболее общая унитарная матрица перехода от перестановочного базиса 
к квантовому --- мы определяем её соотношением 
\Math{\widetilde{U}_j=\transmatr^{-1}P_j\transmatr} --- имеет вид
\Mathh{
\transmatr=\frac{1}{\sqrt{3}}
\Mthree{a}{b~~}{b\runisymb^2}
 {a}{b\runisymb^2}{b~~}
 {a}{b\runisymb}{b\runisymb},
}
где \Math{a} и \Math{b} произвольные элементы множества \Math{\set{1,\runisymb,\runisymb^2}}.
Конкретный выбор этих элементов не играет никакой роли, поскольку в скалярных произведениях,
через которые выражаются наблюдаемые величины, они исчезают, 
входя в произведения сопряжёнными парами.
Поэтому окончательно мы выберем следующую форму для матрицы перехода
\begin{equation}
\transmatr=\frac{1}{\sqrt{3}}
	\Mthree{1}{1}{\runisymb^2}
	 {1}{\runisymb^2}{1}
	 {1}{\runisymb}{\runisymb},~~~~
\transmatr^{-1}=\frac{1}{\sqrt{3}}
	\Mthree{1}{1}{1}
	 {1}{\runisymb}{\runisymb^2}
	 {\runisymb}{1}{\runisymb^2}.
\label{transS3}	 
\end{equation}
Далее мы увидим, что в этой матрице закодирована вся информация о
``квантовом поведении'' группы перестановок \Math{\SymG{3}} с действием на трёх элементах.
\subsubsection{О моделировании квантовых вычислений %с помощью конечных групп
конечными группами}
Реализация квантового алгоритма сводится
к построению унитарного оператора, соответствующего алгоритму, 
из некоторого заданного множества стандартных операторов. 
Существуют универсальные наборы таких операторов. 
\emph{Универсальность} здесь означает, что любой унитарный оператор может быть 
аппроксимирован комбинацией 
стандартных операторов, т.е. эти операторы являются \emph{образующими элементами} 
конечно-порождённой группы, всюду плотной в группе всех унитарных операторов, действующих 
на соответствующем квантовом регистре.
Рассмотрим, например, следующее множество операторов:\\
(a) \emph{оператор Адамара} \Math{H=\frac{1}{\sqrt{2}}\Mtwo{1}{1}{1}{-1}},~
(b) ``\emph{фазовращатель}''	\Math{R\vect{\theta}=\Mtwo{1}{0}{0}{\e^{2\pi{}i\theta}}}
 и (c) ``\emph{контролируемое отрицание}'' 
\Math{\text{CNOT}=\bmat1&0&0&0\\0&1&0&0\\0&0&0&1\\0&0&1&0\emat}.\\
При некоторых значениях параметра \Math{\theta} эти операторы порождают конечную группу.
Например, при \Math{\theta=1/4} на двухкубитном регистре порождается группа \Math{G} 
размера \Math{36864}. Система компьютерной алгебры \textbf{GAP} \cite{gap} 
выдаёт структуру этой группы в таком виде: 
\Mathh{G\cong\vect{\vect{\vect{\vect{\CyclG{8}\times\CyclG{2}}\rtimes\CyclG{2}}
\rtimes\CyclG{3}}\rtimes\CyclG{2}}
\times\vect{\vect{\SL{2}{3}\rtimes\CyclG{4}}\rtimes\CyclG{2}}.}
В случае значений параметра \Math{\theta} в общем положении операторы (a), (b) и (c)
образуют универсальные наборы и, следовательно, порождаемые ими группы бесконечны. 
Однако эти группы в некотором смысле близки к конечным 
--- они являются финитно аппроксимируемыми.
\par
Напомним, что группа \Math{G} 
называется \emph{финитно аппроксимируемой} \cite{Magnus}, 
если для каждого её элемента \Math{g\neq\id} существует такой гомоморфизм 
\Math{\phi: G\rightarrow{}H} 
в \emph{конечную} группу \Math{H}, что \Math{\phi\vect{g}\neq\id}.
Это означает, что любые соотношения между элементами группы \Math{G} 
можно моделировать соотношениями между элементами конечной группы.
Согласно теореме А.И. Мальцева: \emph{всякая конечно порождённая 
группа матриц над полем финитно аппроксимируема}.
\par 
Таким образом, любой универсальный набор операторов 
порождает финитно аппроксимируемую
группу. Это даёт возможность моделировать квантовые вычисления с помощью конечных моделей
по аналогии с широко используемым в физике приёмом, когда для решения некоторой
задачи бесконечное пространство заменяется тором размер которого достаточен для того, 
чтобы вместить данные задачи.

\subsection{Связь математического описания с наблюдением.
%Математическое описание и наблюдение.
\\ Правило Борна}
%\subsection{Правило Борна. Связь математического описания с наблюдением}
Существуют определённые тонкости при переходе от математического описания систем с симметриями
к наблюдаемым ``объектам'' или ``величинам''. 
Подробное обсуждение этой темы можно найти в статье %Г. Вейля 
\cite{Weyl} и в книге \cite{Shafarevich} (стр. 210 и далее). 
Вкратце, дело в том, что для регистрации и отождествления элементов 
системы используются произвольно выбранные метки. 
Объективный смысл имеют только те соотношения и утверждения, которые не зависят от изменений 
в выборе меток поскольку эти изменения представляют собой не более чем переименования.
В системах с симметриями, ``объекты'', составляющие ``однородное'' множество
(более формально, лежащие на одной групповой орбите) имеют различные метки, 
но они неразличимы в абсолютном смысле. Фиксировать такие объекты можно 
только относительно некоторой дополнительной системы, проявляющейся как \emph{система координат} 
или \emph{наблюдатель} или \emph{физический измерительный прибор}. 
Например, невозможно придать абсолютный объективный 
смысл точкам пространства, обозначаемым (помеченным)
 векторами \Math{\mathbf{a}} и \Math{\mathbf{b}}, 
однако отношение между точками, обозначаемое комбинацией 
\Math{\mathbf{b}-\mathbf{a}} 
уже имеет смысл. В более общих групповых обозначения 
эта комбинация может быть записана как \Math{\mathbf{a^{-1}b}}. 
Это пример типичной ситуации, когда наблюдаемые объекты или соотношения
являются групповыми инвариантами, зависящими от \emph{пар элементов}.
Один из элементов такой пары относится к наблюдаемой системе, 
а другой --- к наблюдателю.
\par
В квантовой механике связь между математическим описанием и экспериментом обеспечивается
\emph{правилом Борна} \cite{Landsman}, утверждающим, что \emph{вероятность} наблюдения 
квантовой системы находящейся в состоянии \Math{\psi} аппаратом, настроенным на состояние
\Math{\phi} выражается числом
\begin{equation}
\ProbBorn{\phi}{\psi} = \frac{\textstyle{\cabs{\inner{\phi}{\psi}}^2}}
{\textstyle{\inner{\phi}{\phi}\inner{\psi}{\psi}}}.
\label{Born}	
\end{equation}
Это выражение можно переписать в виде, включающем пару 
``система--аппарат'' более симметрично
\Mathh{\ProbBorn{\phi}{\psi} = 
\frac{\textstyle{\cabs{\inner{\phi}{\psi}}^2}}
{\textstyle{\cabs{\inner{\phi}{\psi}}^2+\Grassnorm{\phi\wedge\psi}^2}}.}
Здесь \Math{\phi\wedge\psi} --- внешнее (Грассманово) произведение векторов 
\Math{\phi} и \Math{\psi}, 
представляющее собой
%\Math{\frac{{\adimH(\adimH-1})}{{2}}}
\Math{\adimH(\adimH-1)/2}-мерный вектор с компонентами в унитарном базисе
\Math{\vect{\phi\wedge\psi}^{ij}=\phi^i\psi^j-\phi^j\psi^i}
и квадратом нормы
\Mathh{\Grassnorm{\phi\wedge\psi}^2=\sum\limits_{i=1}^{\adimH-1}
\sum\limits_{j=i}^{\adimH}\cabs{{\phi^{i}}\psi^{j}-{\phi^{j}}\psi^{i}}^2.}
Обычно в квантовой механике предполагают, что векторы состояний 
нормированы, т. е. \Math{\inner{\phi}{\phi}=\inner{\psi}{\psi}=1}, и записывают правило Борна
в виде \Math{\ProbBorn{\phi}{\psi} = \cabs{\inner{\phi}{\psi}}^2}, что приводит к упрощению
вычислений. Например, легко проверить, что функция \Math{\ProbBorn{\phi}{\psi}} удовлетворяет
основному свойству вероятности --- сумма вероятностей всех возможных результатов наблюдений
состояния \Math{\psi} равна единице. А именно, для любого ортонормального базиса 
 \Math{\set{\onbeq_1,\ldots,\onbeq_\adimH}} в гильбертовом пространстве 
\Math{\Hspace} мы имеем:
\Mathh{\sum\limits_{i=1}^\adimH{}\set{\ProbBorn{\onbeq_i}{\psi}=
\cabs{\inner{\onbeq_i}{\psi}}^2=
\inner{\psi}{\onbeq_i}\inner{\onbeq_i}{\psi}}=
\left\langle\psi\underbrace{\cabs{\sum\limits_{i=1}^\adimH{}
\barket{\onbeq_i}\brabar{\onbeq_i}}}_{\idmat}\emph{}
\psi\right\rangle=\inner{\psi}{\psi}\equiv1.}
Однако мы, стремясь по возможности придерживаться наиболее простых числовых систем, 
не будем использовать нормирование векторов. 
\par
Существуют многочисленные философские спекуляции относительно понятия вероятности 
и её интерпретации, однако на практике в основном используется
\emph{частотная интерпретация}: вероятность --- это отношение числа благоприятных 
случаев к полному числу случаев. Для конечных множеств никаких сложностей 
не возникает вообще --- вероятность это рациональное число, являющееся отношением числа 
выделенных элементов множества к полному числу элементов. 
\par
Можно показать, что
если данные о состояниях системы и аппарата представлены \emph{натуральными числами} 
в перестановочном
базисе, то формула \eqref{Born} даёт \emph{рациональные числа} и в инвариантных
 подпространствах
перестановочного представления, несмотря на то, что промежуточные 
вычисления могут содержать 
циклотомические числа и иррациональности.
\par
Рассмотрим перестановочное действие группы
\Math{\wG=\set{\wg_1,\ldots,\wg_{\wGN}}} на множестве состояний
\Math{\wS=\set{\ws_1,\ldots,\ws_{\wSN}}}. 
Будем описывать состояния системы и аппарата в перестановочном представлении соответственно
векторами 
\begin{equation}
	\barket{n} = \Vthree{n_1}{\vdots}{n_{\wSN}} \text{~и~} 
	\barket{m} = \Vthree{m_1}{\vdots}{m_{\wSN}}.
	\label{natampl}
\end{equation}
Естественно предполагать, что \Math{n_i} и \Math{m_i} --- натуральные числа, интерпретируя их
как ``кратности вхождения'' элемента \Math{\ws_i} в состояния системы и аппарата, соответственно.
Разумеется, ввиду симметрий сами эти числа не наблюдаемы. Наблюдаемыми являются только их инвариантные
%квадратичные 
комбинации. Поскольку стандартное скалярное произведение \eqref{innerstd} 
является инвариантным для перестановочного представления, мы, в соответствии с правилом Борна, имеем
\begin{equation}
	\ProbBorn{m}{n}=\frac{\vect{\sum_i{m_i}n_i}^2}{\sum_i{m_i}^2\sum_i{n_i}^2}.
\label{probp}	
\end{equation}
Ясно, что для ненулевых векторов \Math{n} и \Math{m} с натуральными компонентами
выражение \eqref{probp} представляет собой рациональное число большее нуля, т. е. 
наблюдать, к примеру, деструктивную квантовую интерференцию в такой постановке задачи невозможно в принципе.
\par
Однако деструктивную интерференцию векторов с натуральными компонентами можно наблюдать 
в собственных инвариантных подпространствах перестановочного представления.
Если же вероятности, наблюдаемые в инвариантных подпространствах, ненулевые, то
они представляют собой рациональные числа.
Проиллюстрируем это примером.
\paragraph{Иллюстрация: Группа \Math{\SymG{3}}, действующая на трёх элементах.} 
Векторы состояний в перестановочном базисе имеют вид 
\Math{\barket{n} = \Vthree{n_1}{n_2}{n_3}} и \Math{\barket{m} = \Vthree{m_1}{m_2}{m_3}}.
С помощью унитарной матрицы преобразования \eqref{transS3} можно перевести вектор состояния 
системы \Math{n} из перестановочного базиса в квантовый
\Mathh{
\barket{\widetilde{\psi}}=\transmatr^{-1}\barket{n}
=\frac{1}{\sqrt{3}}
\Mthree{1}{1}{1}
 {1}{\runisymb}{\runisymb^2}
 {\runisymb}{1}{\runisymb^2}\Vthree{n_1}{n_2}{n_3}
=\frac{1}{\sqrt{3}}\Vthree{n_1+n_2+n_3}
{n_1+n_2\runisymb+n_3\runisymb^2}{n_1\runisymb+n_2+n_3\runisymb^2}.
}
Аналогичным образом преобразуется вектор аппарата \Math{m}
\Mathh{
\barket{\widetilde{\phi}}=\transmatr^{-1}\barket{m}
=\frac{1}{\sqrt{3}}\Vthree{m_1+m_2+m_3}
{m_1+m_2\runisymb+m_3\runisymb^2}{m_1\runisymb+m_2+m_3\runisymb^2}.
}
Проекции этих векторов в двумерное представление \eqref{S3umats} имеют вид 
\Mathh{\barket{\psi} = \Vtwo{n_1+n_2\runisymb+n_3\runisymb^2}
{n_1\runisymb+n_2+n_3\runisymb^2},~~~~ 
\barket{\phi} = \Vtwo{m_1+m_2\runisymb+m_3\runisymb^2}
{m_1\runisymb+m_2+m_3\runisymb^2}.} Мы отбросили здесь коэффициент \Math{1/\sqrt{3}}
поскольку борновская вероятность --- проективный инвариант. 
Заметим, что векторы \Math{\psi} и \Math{\phi} исчезают тогда и только тогда, когда
\begin{equation}
n_1=n_2=n_3 \text{~~и~~} m_1=m_2=m_3,
	\label{zerocond} 
\end{equation}
поскольку примитивный корень единицы \Math{\runisymb} в данном случае удовлетворяет
соотношению \Math{1+\runisymb+\runisymb^2 = 0}. 
Условия \eqref{zerocond} определяют 
собственный вектор одномерного тривиального подпредставления, 
ортогонального рассматриваемому двумерному.
\par
Для двумерной подсистемы, выражения, входящие в формулу 
\eqref{Born} борновской вероятности, имеет вид
\begin{equation}
	\inner{\psi}{\psi}=3\vect{n_1^2+n_2^2+n_3^2}-\vect{n_1+n_2+n_3}^2,
\label{Born2den1}	
\end{equation}
\begin{equation}
	\inner{\phi}{\phi}=3\vect{m_1^2+m_2^2+m_3^2}-\vect{m_1+m_2+m_3}^2,
\label{Born2den2}	
\end{equation}
\begin{equation}
\cabs{\inner{\phi}{\psi}}^2=\vect{3\vect{m_1n_1+m_2n_2+m_3n_3}
-\vect{m_1+m_2+m_3}\vect{n_1+n_2+n_3}}^2.
\label{Born2num}	
\end{equation}
Заметим, что 
\begin{enumerate}
	\item Выражения \eqref{Born2den1}--\eqref{Born2num} состоят из \emph{инвариантов
перестановочного представления}. 
Это подчёркивает фундаментальную роль перестановок в квантовом описании.
	\item Выражения \eqref{Born2den1} и \eqref{Born2den2} всегда положительные целые числа.
	(Если только не выполняются условия \eqref{zerocond}, при которых эти выражения обращаются в нуль.)
	\item	Условия \emph{деструктивной квантовой интерференции}, 
	т.е. обращение в нуль борновской
	вероятности \eqref{Born}, определяются уравнением
\Mathh{3\vect{m_1n_1+m_2n_2+m_3n_3}-\vect{m_1+m_2+m_3}\vect{n_1+n_2+n_3}=0.}
Это уравнение имеет бесконечное множество решений в натуральных числах.
Пример такого решения: 
\Math{%\set{
\barket{n} = \Vthree{1}{1}{2},~~\barket{m} = \Vthree{1}{3}{2}}.
\end{enumerate}
Таким образом, мы, простым переходом к инвариантному подпространству, получили 
существенные черты квантового поведение из ``перестановочной динамики'' 
и ``натуральной'' интерпретации \eqref{natampl} квантовой амплитуды.
\section{Заключение} 
Анализ квантового поведения с помощью конечных моделей приводит к выводу, что квантовая
механика не столько физическая теория, сколько в достаточной степени априорная 
математическая схема в основе которой лежит неразличимость объектов --- своего рода 
``\emph{исчисление неразличимых}''
(по аналогии с термином непрерывной математики ``\emph{исчисление бесконечно малых}'').
В основе квантового поведения лежит фундаментальная невозможность проследить 
тождественность однородных объектов в процессе их эволюции. 
\par
Г. Вейль по этому поводу писал \cite{Weyl} следующим образом: 
``\emph{В настоящее время мы говорим только о том, сколько элементов \Math{n_i\vect{t}}
находится в состоянии \Math{C_i\vect{t}} в любой момент \Math{t}, поскольку мы не можем
проследить тождественность \Math{n} индивидов во времени. Мы не знаем про элемент, 
находящийся в данный момент, например, в состоянии \Math{C_5} был ли он в предыдущий
момент в состоянии \Math{C_2} или \Math{C_6}.}''
\par
Поэтому объективными могут быть только (``статистические'') утверждения о числах
некоторых инвариантных комбинаций элементов. Эти утверждения должны выражаться в 
терминах групповых инвариантов и натуральных чисел 
(необязательно взаимно независимых), характеризующих группы симметрий, 
таких как размеры орбит, размеры классов сопряжённых элементов, 
коэффициенты алгебры классов, размерности неприводимых представлений, и т.д.
\paragraph{Благодарности.}
Работа частично финансировалась за счёт грантов 
№ 01-01-00200 Российского Фонда Фундаментальных Исследований
и НШ-3810.2010.2 Министерства образования и науки Российской Федерации.

\end{document}